# An aspherical distribution for the explosive burning ash of core-collapse supernovae

Qiliang Fang[1]*, Keiichi Maeda[1]*, Hanindyo Kuncarayakti[2,3], Takashi Nagao[2,4,5]

[1]Department of Astronomy, Kyoto University; Kyoto, 606-8502, Japan.

[2]Tuorla Observatory, Department of Physics and Astronomy, FI-20014 University of Turku, Finland.

[3]Finnish Centre for Astronomy with ESO (FINCA), FI-20014 University of Turku, Finland.

[4]Aalto University Metsähovi Radio Observatory, Metsähovintie 114, 02540 Kylmälä, Finland

[5]Aalto University Department of Electronics and Nanoengineering, P.O. BOX 15500, FI-00076 AALTO, Finland

*Corresponding authors. Email: fangql@kusastro.kyoto-u.ac.jp; keiichi.maeda@kusastro.kyoto-u.ac.jp

**It is widely believed that asphericity in the explosion is the crucial ingredient leading to successful core-collapse supernovae (CCSNe). However, the direct observational evidence of the explosion geometry, and its connection with the progenitor properties, are still missing. Based on the thus-far largest late-phase spectroscopy sample of stripped-envelope CCSNe, we demonstrate that about half of the explosions exhibit substantial deviation from sphericity. For these aspherical CCSNe, the spatial distributions of the oxygen burning ash and the unburnt oxygen, traced by the profiles of [Ca II] λλ7291,7323 and [O I] λλ6300,6363 emissions respectively, appear to be anti-correlated, which can be explained if the explosion is bipolar in nature and the oxygen-rich material are burnt into two detached iron-rich bubbles. The combined analysis of the explosion geometry and the progenitor mass further suggests the degree of asphericity grows with the carbon-oxygen core, which holds the promise to guide state-of-the-art simulations of CCSN explosion.**

When the nuclear fuel deep inside a massive star (with the zero-age main-sequence mass > 8 $M_{\odot}$) is exhausted, the core collapses to form a neutron star or a black hole. A large amount of the gravitational energy is released, part of which is transformed to the kinetic energy expelling the rest of the star, leading to a core-collapse supernova (CCSN).

The mechanism behind CCSN is one of the major long-standing challenges in astrophysics. The geometry of the explosion, and its relation with the mass of the progenitor, should shed light on this unresolved issue. If the released energy is concentrated in a specific direction, the explosion occurs in a bipolar configuration (*1,2*). During this process, the oxygen-rich materials are compressed to the equatorial plane, whereas the explosive burning takes place along the axis perpendicular to the oxygen-rich plane (*3,4*). The bipolar explosion is of particular interest due to its potential connection with gamma-ray bursts (GRBs), which requires a collimated relativistic beam and can be observed only when viewed on-axis (*5*).

The late-phase (nebular) spectroscopy of stripped-envelope supernovae (SESNe, a subtype of CCSNe), is a unique tool to study the geometry of the innermost ejecta. An SESN is produced by a massive star that has lost most of its hydrogen envelope before the explosion (*6*). Several months after the explosion, without being blocked by the massive envelope, the ejecta of an SESN becomes optically thin following its expansion, and the central region is exposed. During this phase, the spectrum of an SESN is dominated by forbidden lines, in particular [O I] λλ6300,6363 and [Ca II] λλ7291,7323 (Fig. 1). Research works in the past focused on [O I] emitted from the unburnt oxygen-rich stellar material, the profile of which strongly depends on the viewing angle if the explosion is aspherical (*7,8*). The double-peaked [O I], a distinct feature of a bipolar explosion, has been found to be common for SESNe (*9*). This result has been further justified based on larger samples (*10-13*). However, inferring the geometry of the explosion using only [O I] falls into a dilemma: (1) When viewed from the on-axis direction, [O I] will appear to be sharp and single-peaked, making it indistinguishable from a spherical explosion; (2) the double-peaked [O I], which is a smoking gun of a bipolar explosion, suggests the explosion is viewed on-edge, then its possible connection with a GRB would be missing (*7*); (3) Further, the interpretation of a two-dimensional configuration based on a one-dimensional line profile is degenerated (*11-13*). We therefore need an improved method to infer the explosion geometry, especially the characteristic features when the bipolar explosion is viewed on-axis.

When a CCSN explosion occurs, oxygen is burnt into iron-peak elements in the direction in which a large amount of the energy is released. Outside of this region, the material is still oxygen-rich, and the fractions of the heavy elements are negligible. The explosive burning ash and the unburnt material are therefore decoupled in geometry. The relation of the profiles of the emission lines originating in these two regions is thus a good probe to the explosion geometry. [O I] has been frequently employed to infer the geometry of the unburnt oxygen-rich material, while [Ca II], which is strong in most SESNe, is the main coolant of the oxygen burning ash, making it an ideal tool to trace the configuration of the explosion-made region (*14-18*).

In this work, we propose to combine the analyses of the [O I] and [Ca II] profiles to infer the geometry of the explosion. Horn-like and flat-topped profiles in either of the two lines are considered as the signature of an aspherical explosion. Hereafter they are collectively referred to as double-peaked (DP) profiles, to distinguish from single-peaked (SP) profiles. We look into our sample of nebular spectra of 80 SESNe. Among these objects, 41 show single-peaked [O I] and single-peaked [Ca II] (OSCaS). For objects considered as bipolar explosions (i.e., with either double-peaked [O I] or [Ca II]), the [O I] and [Ca II] profiles appear to be anti-correlated: 27 objects have double-peaked [O I] and single-peaked [Ca II] (ODCaS) while 12 objects have single-peaked [O I] and double-peaked [Ca II] (OSCaD). Some examples of SNe fall into different line profile categories are shown in Fig. 1. The anti-correlation of the [O I] and [Ca II] profiles can be understood as an outcome of different viewing angles to observe a similar explosion configuration, namely the bipolar explosion which is characterized by the burning ash distributed along a specific axis, surrounded by an oxygen-rich torus. When viewed on-axis, the Doppler shifts of the bipolar calcium elements give rise to the horn-like profile of [Ca II], while most of the oxygen elements on the torus move perpendicularly to the line of sight (LOS), producing single-peaked [O I]. On the other hand, when viewed from the direction perpendicular to the axis, [O I] appears to be horn-like and [Ca II] is single-peaked for a similar reason. This simple picture is supported further by the data; (1) the line profiles are largely either single-peaked or double-peaked, and (2) there is no object showing both horn-like [O I] and [Ca II].

Evidence of asphericity can also be inferred from the widths of [O I] and [Ca II]. Since the velocity of the material is proportional to the radial coordinate under the assumption of homologous expansion, the width of a line reflects the spatial extension of the emitting elements along the LOS (*11,13,14,16,18*). The full-widths at half-maximum (FWHM) of [O I] and [Ca II] are compared in Fig. 2A. For the OSCaS and ODCaS objects, [O I] is broader than [Ca II] in more than 80% of cases. While for OSCaD objects, the [Ca II] is broader than [O I] in most cases. The difference between the [Ca II] and [O I] widths is shown in Fig. 2B. The OSCaD objects have distinct population from the other types, with $p < 0.001$ from the Anderson-Darling (AD) test. The large FWHM of the [Ca II] contradicts with a spherical explosion, where heavier elements are explosively synthesized in inner regions and expelled at lower velocities. However, it can be naturally explained by the bipolar explosion; heavy elements are synthesized along a specific axis and can be ejected at high velocities. If viewed closely from the axis, the average LOS velocity of the iron-rich ash can be faster than that of the oxygen-rich material, resulting in broader [Ca II] (*4*).

The peculiar double-peaked profile of the [Ca II], along with its broadness, strongly suggests a bipolar distribution in its emitting region. We propose an idealized axisymmetric model to explain the profiles of the OSCaD and ODCaS objects (Fig. 3). The model is characterized by two separated zones: the calcium elements are distributed in a peanut-like region ($X_{\rm O}$ =0, $X_{\rm Ca}$=1. $X$ is the mass fraction), while outside of this region, the material is oxygen-rich ($X_{\rm O}$ =1, $X_{\rm Ca}$=0). The physical properties (e.g., mass fractions, gamma-ray deposition rate) are assumed to be uniform in each zone for simplicity. No radiation transfer process is included. These simplifications allow for focus on the geometric effect. When viewed from $\theta = 15^{\circ}$, 50º and 75º from the axis, the model produces single-peaked [O I]/horn-like [Ca II], single-peaked [O I]/flat-topped [Ca II] and horn-like [O I]/single-peaked [Ca II] respectively, in good agreement with the observed profiles of SNe 2008im, 2005nb and 2004ao shown in Fig. 3 as examples. It is surprising that such a simple model involving only 2 parameters (the maximum velocity $V_{\rm max}$ and the viewing angle $\theta$) can explain the wide range of both the [O I] and [Ca II] profiles simultaneously, as well as their anti-correlation. This result suggests that the geometric effect is responsible for the diversity seen in the line profiles. Including other ingredients (e.g., the density distribution, radiation transfer) may improve the alignment with the observation, although their roles are probably secondary.

Statistics is useful to constrain the explosion geometry. The occurrence rate of bipolar explosions ($N_{\rm Bipolar}$=$N_{\rm DP\ [Ca\ II]}$ + $N_{\rm DP\ [O\ I]}$; Here $N_{\rm DP\ [Ca\ II]}$ and $N_{\rm DP\ [O\ I]}$are the numbers of objects showing double-peaked [O I] and double-peaked [Ca II], respectively) is 0.49 (= 39/80), implying about half of the SESNe explode with a non-negligible departure from a spherical configuration. Within this bipolar explosion category, the occurrence rate of double-peaked [Ca II] is 0.31 (= 12/39), which is consistent with the axisymmetric model: at $\theta \sim 45^{\circ}$, the [Ca II] profile transforms from double-peaked to single-peaked. If no bias in orientation is involved , the expected occurrence rate of double-peaked [Ca II] in this model is $0.29 \pm 0.06$ for this sample size, assuming that the solid angle is uniformly distributed.

The dependency of the explosion geometry on the progenitor mass is crucial to reveal the explosion mechanism of CCSNe. However, observational hints are still missing. In this work, we employ the intensity ratio of [O I] to [Ca II] (hereafter denoted as [O I]/[Ca II]) as a proxy of the CO core mass under the assumption that the amount of the burning ash is not sensitively dependent on the CO core mass. This relation is routinely adopted by past theoretical and observational works, with a larger [O I]/[Ca II] implying more massive CO core (*13-15,19-23*). To investigate the relation between [O I]/[Ca II] and the asphericity of the explosion, the sample is binned into 10 groups according to the [O I]/[Ca II] sequence. The range of [O I]/[Ca II] of the bins are -0.3 to 0.6 in logarithmic scale, which roughly corresponds to the CO core mass of 2.05 to 7.84 $M_{\odot}$, or the ZAMS mass of 12 to 30 $M_{\odot}$, based on the scaling relations between these quantities (*20,24*). The relation between [O I]/[Ca II] and the bipolar rate ($N_{\mathrm{DP\ [Ca\ II]}} + N_{\mathrm{DP\ [O\ I]}}$ divided by the total number of the objects) in each group is investigated: no correlation is discerned (Spearman correlation coefficient $\rho = 0.03$), as shown in Fig. 4A.

The lack of correlation between the bipolar rate and [O I]/[Ca II] possibly suggests that whether a CCSNe explodes in a spherical or bipolar configuration is not determined by its CO core mass. However, the association of [O I]/[Ca II] to the CO core mass will require further quantification; for example, if the explosion energy tends to be larger for a more massive CO core (*3,13,24,25*), creating a larger amount of Ca (*26,27*), this relation may be diluted. In any case, the statistical investigation presented here indicates that a common explosion mechanism is behind CCSNe across a range of the CO core mass, at least in the sample of the canonical SESNe studied in the current work; the asphericity in the energy release must be a key to producing CCSNe, with about half of the sample showing the detectable level of the bipolar distribution as seen either in [O I] or [Ca II].

A similar practice is separately applied to the bipolar sample. The cumulative distributions of [O I]/[Ca II] of the OSCaD and ODCaS objects are compared with the full sample in Fig. 4B. Surprisingly, the objects with double-peaked [Ca II] have a larger [O I]/[Ca II] ratio than those with double-peaked [O I] ($p < 0.05$ from the AD test). At the low [O I]/[Ca II] range, the presence of double-peaked [O I] suggests that the explosion is bipolar, while the iron-rich bubbles that characterize the geometry of the burning ash are not distinctly separated in two hemispheres, as inferred from the absence of double-peaked [Ca II]. By binning the bipolar SNe into 5 groups according to the [O I]/[Ca II] sequence, we find a trend that the occurrence rate of the double-peaked [Ca II] in each group increases with the average [O I]/[Ca II] ($\rho = 0.87$, $p = 0.053$; See Fig. 4C), although the statistics is limited by the relatively small number in each bin ($N \approx 8$) and should be considered as suggestive (*28*). This tendency can be explained if the transition angle $\theta_{\mathrm{trans}}$ (above this angle the [Ca II] profile transforms from double-peaked to single-peaked) is increasing with [O I]/[Ca II]. This is consistent with the scenario where the explosion becomes more aspherical, and the burning ash becomes more detached and more collimated, as the CO core grows, which is schematically demonstrated in Fig. 4C.

The correlation of the occurrence rate of double-peaked [Ca II] (Fig. 4C) and the probable non-correlation of the occurrence rate of the aspherical explosions (Fig. 4A) to the CO core mass (or explosion energy, considering its correlation with the kinetic energy $E_{\mathrm{K}}$ of the ejecta; *Refs. 13*,*24*) must be satisfied in the explosion mechanism(s). The standard delayed-neutrino explosion mechanism involves various instabilities (*29*), some of which may introduce a bipolar configuration (e.g., standing accretion shock instabilities or SASI; *Refs.30,31*). Recent simulations further show that either rotation or magnetic field, or both, may have an important role to assist the neutrino-driven explosion (*32,33*); these effects may naturally introduce a specific direction in the explosion dynamics. In a massive core, the neutrino-driven mechanism alone would be insufficient to overcome the binding energy (*29*), and then the rotation and magnetic field may become important even energetically, i.e., the magneto-rotational explosion (*34)*.

# Methods

## Sample description

The SESN sample in this work is the same with the one used in reference (*13*), most of which are compiled from the literature (*35,36*). The analysis is restricted to the nebular spectra acquired not less than 100 days after the discovery of the SN. In this sample, 80 objects have spectra with a high signal-to-noise ratio to allow the investigation of both the [O I] λλ6300,6363 and [Ca II] λλ7291,7323 profiles, and are selected for further analysis. SNe associated with a GRB or X-ray flash (XRF, a low energy analog of GRB), SNe 1998bw/GRB 980425, and 2006aj/XRF 060218, are already excluded from the analysis to avoid introducing any bias in orientation (*37-40*). SN 2012ap, which is suggested as a GRB-associated SN but viewed off-axis, is also excluded (*41,42*). The peculiar SN 2005bf, whose explosion mechanism is probably different from normal SESNe, is not included either (*9,43*). The SESNe in this work are listed in Supplementary Tab. 1.

## Line width measurement and error estimation

Before measuring the line widths of [O I] and [Ca II], the nebular spectrum is firstly corrected for redshift, which is derived from the central wavelengths of the narrow emission lines from the explosion site or the redshift of the host galaxy (*44*). The next step is to remove the underlying continuum following the same method as previous works (*13,22,45*), where the line connecting the minima at both sides of the emission is defined to be the local continuum and then excised, as illustrated in Supplementary Fig. 1.

To derive the isolated [O I], we need to subtract the broad Hα-like structure located at the red wing of [O I]. This emission is assumed to be symmetric with respect to 6563 Å, and is constructed by reflecting its red side to the blue side. A similar practice is applied to derive the [Ca II] profile, which is complicated by many emissions at similar wavelength ranges emitted by the iron-peak elements. However, most of these lines are usually very weak compared with [Ca II], and is not taken into consideration to avoid further complication. We first subtract He I λ7065 line by assuming it is symmetric with respect to its central wavelength, and the profile is constructed by reflecting its blue side to the red side. After He I is subtracted, the profile of [Fe I] λ7155 is constructed based on the same method. The separation of the central wavelengths of [Fe I] and [Ca II] is ~5700 km $s^{-1}$, while the half-width at half maximum (HWHM) of [Ca II] is always smaller than ~5000 km $s^{-1}$ (see Fig. 2 in the main text). The blue side of [Fe I] is therefore hardly contaminated by [Ca II]. [Ni II] λ7377 is difficult to remove from the [Ca II] complex, because its separation from the central wavelength of [Ca II] is small (~3500 km $s^{-1}$). We assume that [Ni II] and [Fe I] have the same profile, and the ratio of the intensities $L_{7377}/L_{7155}$ is 0.6 (for the solar Ni/Fe ratio and temperature of the emission region ~4000K; see *46,47*). The profile of [Ca II] is derived by subtracting the He I, [Fe I] and [Ni II] profiles from the complex. Other contaminating lines are usually very weak compared with [Ca II], and they are not taken into consideration to avoid further complication. The FWHM of the isolated [O I] and [Ca II] are then measured and corrected for the instrumental broadening. Some examples of the FWHM measurement are shown in Supplementary Fig. 1.

The uncertainty of the FWHM is estimated by employing Monte Carlo method. The original spectrum is smoothed by convolving it with a boxcar filter. The smoothed spectrum is subtracted from the original one, then the standard deviation of the residual flux at the wavelength range 6000 to 8000 Å is estimated as the noise level of the spectrum. To generate a simulated spectrum, Gaussian noise is added to the spectrum using the estimated noise level. The end points of the local continuum, the central wavelengths of the Hα-like structure, He I, [Fe I] and [Ni II] are allowed to varied by ± 600 km $s^{-1}$ to account for the typical resolution of nebular spectroscopy, assuming flat distributions. The FWHM is then calculated by following the method described above. This procedure is repeated for $10^4$ times, and the 16 and 84 percentiles of these measurements are estimated as the lower and upper limits of the FWHM.

## Line morphology

The profiles of [O I] and [Ca II], being diverse in morphology, are employed to characterize the geometry of the emitting regions. While the [O I] profile is typically categorized by multi-Gaussian fitting, this method has several limitations: (1) the results of the fitting can be degenerated, leading to ambiguous interpretation of the profile; (2) the blending of the emissions from iron-peak elements complicates the extension of this approach to [Ca II]. In this work, we employ the convolution method (*48*). The line profiles of [O I] in SN 2008im and [Ca II] in SN 2011dh are employed as the templates of single-peaked [O I] and [Ca II], respectively. These templates are convolved with Gaussian kernels by freely adjusting the centers and standard deviations until their widths match with those of the other objects to ensure proper comparison. Profiles that resemble with the convolved template are classified as single-peaked (SP). Alternatively, profiles that deviate from the convolved template, but exhibit troughs located at ~ 0 km $s^{-1}$ and roughly symmetrical local peaks with respect to 0 km $s^{-1}$ fall into the horn-like category. In the absence of these features, the profiles are classified as flat-topped. It should be noted that the flat-topped and horn-like profile sometimes cannot be unambiguously classified, especially when the troughs are shallow. However, the exact distinction between these two profiles is not essential to the analysis in this work. Therefore, the horn-like and flat-topped profiles are collectively referred to as double-peaked (DP) profiles. We neglect the substructures that occasionally emerge on the broad bases of the emissions and focus on the bulk properties. In practice, we find this evaluation leads to unambiguous classification between SP and DP profiles. Several examples are demonstrated in Supplementary Fig. 2.

## Axisymmetric model

Rather than working on multi-dimensional radiative transfer with detailed treatment of the hydrodynamics and nucleosynthesis, we attempt to develop a simplified axisymmetric model that can reproduce the unique double-peaked profiles of [O I] and [Ca II], and their anti-correlation. The axisymmetric model is characterized by two separated zones: (1) the explosion burning ash described by two symmetric (with respect to the plane perpendicular to the axis of symmetry), off-center ellipsoids. The velocity of the edge of the ellipsoid is $V_{\rm max}$, the maximum velocity of the model. In this zone, the oxygen elements synthesized during stellar evolution are burnt into heavier elements, with $X_{\rm O}$ = 0 and $X_{\rm Ca}$ = 1 (we are only interested in calcium, other iron-peak elements are thus not included); (2) the O-rich zone enclosed in the sphere with radius $V_{\rm O} = 0.8V_{\rm max}$. We fix $V_{\rm O} = 0.8V_{\rm max}$ ,

as this value can well produce the widths of both [O I] and [Ca II] for most SNe in the sample except for 2 objects, SNe 2006T and 2012fh. The overlap of this sphere with the iron-rich ellipsoids is excised. This zone is filled with the unburnt material where the fractions of the iron-peak elements are negligible ($X_O$ = 1 and $X_{Ca}$ = 0). The mass fractions and the density are assumed to be homogeneous. Throughout the model construction, we use velocity as the proxy of spatial coordinate: during nebular phase, the ejecta dynamics is well characterized by homologous expansion, where a grid point is moving along the radial direction with velocity proportional to its radial coordinate.

The above model is mapped onto a 3-dimensional cube, which is divided into 500×500×500 grids in the Cartesian coordinate, with each side bounded by $-V_{max}...+V_{max}$. No radiation transfer is introduced. The flux from each grid point is assumed to be proportional to its volume. Homologous expansion suggests that the infinitesimal volume $dr^3$ is proportional to its "volume" $dV^3$ in velocity space. The specific flux of [O I] (or [Ca II]) $F_V dV$ at velocity $V$ is calculated by summing up $dV^3$ of the O-rich (or iron-rich) grid points with the LOS velocities within $V...V+dV$. [O I] is in nature doublet, consisting of two transitions emitting at 6300 and 6363 Å respectively. These two components, separated by ~3000 km $s^{-1}$, have intensity ratio 3:1 in the optically thin limit. For [Ca II] doublet, the two components, centered at 7291 and 7323 Å, are separated by ~1300 km $s^{-1}$, and their intensity ratio is fixed to be 1:1 in the optically thin limit. The model [O I] and [Ca II] profiles, with $V_{max}$ = 5000 km $s^{-1}$ and viewing angle $\theta$ varies from 0° (on-axis) to 90° (on-edge), are shown in Supplementary Fig. 3.

The axisymmetric model predicts the unique features of both [O I] and [Ca II] profiles, and their anti correlation. The [O I] and [Ca II] profiles are determined by $\theta$ with the opposite tendency: when viewed from $\theta$ = 0° (on-axis), the model [Ca II] shows a horn-like profile and [O I] is single-peaked; when viewed from $\theta$ = 90° (on-edge), the [Ca II] profile becomes single-peaked and [O I] transforms to the horn-like profile.

## Application of the model to the observation

In this section, the model profiles are applied to SNe with either double-peaked [Ca II] or [O I] (except for SN 2015ah which will be discussed later). We freely vary $V_{max}$ and $\theta$ to find the parameters that correspond to the observation. The procedure is as follow: for each object, we focus on the line that exhibits the double-peaked feature, i.e., [O I] for ODCaS category and [Ca II] for OSCaD category. The viewing angle of the model is varied until the depth of the trough in the corresponding line matches the observation. Flat-topped profiles can be viewed as the extreme case that the trough depth is 0. After $\theta$ is fixed, we adjust $V_{max}$ until the widths of the observed DP profile and the corresponding model profile match. The line that appears to be SP is not directly involved in this procedure, but serves to assess the alignment between the model and observation. In two cases, SNe 2006T and 2012ah, the third parameter, i.e., the maximum velocity of the O-rich region $V_O$, is involved, although in most cases fixing $r_O$ (= $V_O/V_{max}$) = 0.8 already produces reasonable matches. The applications of the model to the full sample of SNe that are considered as bipolar explosions are demonstrated in Supplementary Fig. 4.

The [O I] and [Ca II] profiles from the simplified model indeed captures the main features of [O I]

and [Ca II], although failed to produce some substructures, which may require introducing clumpy structures (for example, a moving O-rich blob as exemplified by SN 1996aq; See *Ref. 11*) or a centrally-concentrated density profile (SN 1990aj, SN 2002ap and SN 2006ck). For the [O I] profiles being horn-like, the model predicts that the intensity of the peak at the redder wavelength is larger than the bluer one because of the contribution from the secondary component of the [O I] doublet centered at 6363 Å (or 3000 km $s^{-1}$). However, some objects shows flat-topped profiles (e.g., SN 1990U and SN 2005aj) or a stronger blue-peak (SN 2000ew, SN 2004gt and iPTF13bvn). This could be explained by the effect of the residual opacity: a photon escaped from the far side of the ejecta experiences twice the scattering and/or absorption (see for example *Ref. 20*), which will effectively reduce the fluxes at redder wavelengths. If the fluxes from the grid points with LOS velocity > 0 are manually reduced by 10-40%, the model indeed produces [O I] with the flat-topped profile or horn-like profile with the stronger blue-peak.

In conclusion, the axisymmetric model can produce the [O I] and [Ca II] profiles of the SESNe with signatures of a bipolar explosion in the sample of this work, as well as their anti-correlation. Including other ingredients (clumpy structure, centrally concentrated density profile or non-axisymmetry) and radiative transfer effects will better capture the substructures of the profiles, but the bulk properties, i.e., single peak versus double peak, will not be substantially affected.

## Connection with gamma-ray bursts

Although we have excluded SNe associated with GRB from our analysis in the main text, the following events are closely examined in this section. SN 1998bw, associated with GRB 980425, has been suggested to be viewed on-axis, while SN 2012ap has been suggested to be associated with a GRB but viewed off-axis (*37,38*,*41*,*42*). The prior knowledge on the orientation of these two objects allows us to employ them as test cases to evaluate the prediction of the model in this work.

The [O I] and [Ca II] profiles of SNe 1998bw and 2012ap are shown in Supplementary Fig. 5. The spectra are observed on 1998 Nov 26$^{th}$ (220 days after the explosion; *49*) and 2012 Nov 16$^{th}$ (272 days after the explosion; *41*) respectively. The [O I] profile of SN 1998bw appears to be single-peaked, while that of SN 2012ap exhibit a trough at ~ 0 km $s^{-1}$, and is therefore classified as double-peaked. Although the [O I] profiles of both objects aligned with model prediction, their [Ca II] profiles exhibit peculiar characteristics. In the case of SN 1998bw, [Ca II] appears to be flat-topped, with a trough centered at -2500 km $s^{-1}$ identified. The overall shape resembles to the [Ca II] profile of SN 2002ap, a prototype of flat-topped profile, except for the large blueshift. If the trough is viewed as the center of [Ca II], the explosion of SN 1998bw must be highly aspherical, which cannot be explained by the axisymmetric model in this work: in any case, the trough of the double-peaked [Ca II] is predicted to be located at ~0 km $s^{-1}$. The lack of [Ca II] flux at redder wing suggests that the burning ash is either highly opaque, trapping the photons from the rear side, or is uni-polar in nature (*4*). Additionally, the [Ca II] complex of SN 1998bw is also found to be blended by strong [Fe I] λ7155 at its blue wing, which is typically very weak in normal SESNe. These factors complicate the backward inference of the geometry of the explosion-made region from [Ca II] profile. Regarding SN 2012ap, its [Ca II] is single-peaked, but much broader than [O I]. A possible explanation for this broadening is contamination by the emissions from the iron-peak elements. However, [Ca II] of this object is too broad to allow the de-blending procedure employed in this work.

To illustrate these points, the model profiles are compared with SN 1998bw and 2012ap. By fixing the viewing angle to be $\theta = 0^{\circ}$ (SN 1998bw) and $\theta = 75^{\circ}$ (SN 2012ap), we adjust $V_{max}$ to match the widths of observed and model [O I]. While the [O I] profiles of both objects are well reproduced, the model [Ca II] fail to capture the main features of the observed [Ca II], as illustrated in Supplementary Fig. 5. For SN 2012ap, if we assume Fe and Ni follow the same spatial distribution with Ca, and adjust the intensity ratios of the lines ($L_{7155}/L_{7291}$~0.56, $L_{7377}/L_{7291}$~0.12 and $L_{7412}/L_{7291}$~0.31), the [Ca II] observed complex can be reproduced to some extent. It should be noted that the model profiles are blueshifted by 1000 km $s^{-1}$ to match with the peak of the observed profile. However, introducing these emissions can not explain the peculiar [Ca II] profile of SN 1998bw: in any case the model profiles are too broad.

The failure in reproducing the [Ca II] profile of SN 1998bw suggests that the axisymmetric model may be overly simplified to explain the explosion geometry of GRB-associated SNe. Further, the unusually abundant iron-peak elements seen in both objects indicate notable differences in explosive nucleosynthesis compared to normal SESNe. These observations, coupled with the absence of GRB association in normal SESNe even when they are viewed on-axis, suggests that the bipolar explosion alone may not be sufficient to generate GRB: more extreme conditions are probably required.

## Peculiar object: SN 2015ah

SN 2015ah is classified as an OSCaD object, with the [Ca II] complex showing the horn-like profile with trough located at ~ 4000 km $s^{-1}$, similar to the central wavelength of [Ni II] λ7377 (~ 3600 km $s^{-1}$). There is also a shallow trough located at the central wavelength of [Fe II] λ7155 (~ -5600 km $s^{-1}$). These features suggest this complex is probably not dominated by [Ca II] as the rest of the sample. Instead, [Ni II] may be the main contributor. Both the [Fe I] and [Ni II] lines are emitted from the burning ash. Assuming that they share the same spatial distribution as the [Ca II] emitting region, by adjusting the intensity ratios of the lines ($L_{7377}/L_{7291}$ ~ 3.7 and $L_{7155}/L_{7291}$ ~ 1.1), the model indeed reproduces the profile of the [Ca II] complex, as shown in Supplementary Fig. 6.

The strong [Ni II] ($L_{7377}/L_{7155}$ ~ 3.4) suggests this object exhibits substantial over production in the Ni/Fe ratio (*46,47*). To our knowledge, this is the third CCSNe with such a super-solar Ni/Fe ratio reported. The other two cases are SNe 2012ec (*47*) and 2006aj (*50*). Although the sample is still very small, it seems that the excess in the Ni/Fe ratio can occur for all types of CCSN, irrespective of the presence or absence of the outer envelope. Further, two out of the three cases, SNe 2006aj and 2015ah, are likely associated with non-sphercial explosions, as the asymmetry in explosion can effectively eject the deep-lying burning ash, leading to the change in the Ni/Fe ratio (*3*). A systematic investigation of the relations between the Ni/Fe ratio and the properties of the explosion will be important to reveal the origin of the iron-peak elements, and provide constraints on the explosion mechanism of CCSNe.

**Data availability:** Most of the spectra are available from WiseRep (https://www.wiserep.org/) and Supernova Database of Berkeley (http://heracles.astro.berkeley.edu/sndb/). The data that support the plots within this paper and other findings of this study are available from the corresponding author upon reasonable request.

**Code availability:** Astropy, Matplotlib, Numpy and Scipy are available from the Python Package Index (PyPI) (https://pypi.org/). Upon request, the first author will provide the python code used to generate the model line profiles.

**Acknowledgments:** The authors thank Ji-an Jiang, Seppo Mattila and Anders Jerkstrand for their reviewing the pre-submission manuscript and providing a number of constructive suggestions. The authors thank the referee for their constructive suggestions. Q.F. acknowledges support from the Japan Society for the Promotion of Science (JSPS) KAKENHI grant 20J23342. K.M. acknowledges support from the JSPS KAKENHI grant JP18H05223, JP20H00174, and JP20H04737. H.K. and T.N. are funded by the Research Council of Finland projects 324504, 328898 and 353019. This work is supported by the JSPS Open Partnership Bilateral Joint Research Projects between Japan and Finland (K.M and H.K; JPJSBP120229923).

**Author contributions:** Q.F., K.M. and T.N. initialized the project. Q.F. led the nebular spectroscopy analysis, model construction and the manuscript preparation. K.M. organized the efforts for interpretation of the results and assisted in manuscript preparation. H.K. and T.N. contributed to the spectroscopy analysis and interpretations. All authors contributed to discussion and editing the manuscript.

**Competing interests:** The authors declare no competing interests.

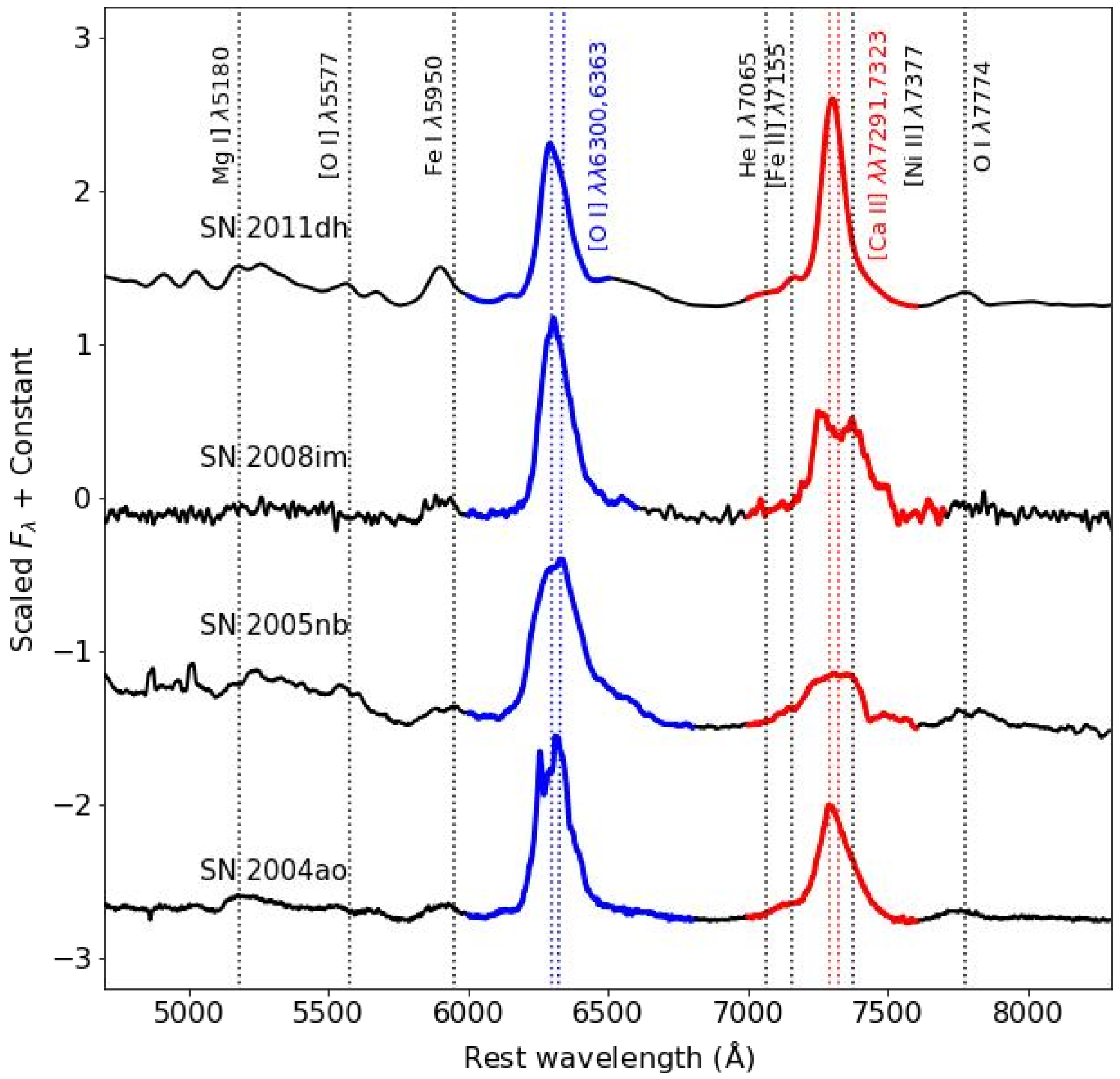

**Fig. 1 Nebular spectroscopy of SESNe.** The ranges of [O I] (blue) and [Ca II] (red) are highlighted. From top to bottom: SNe 2011dh (single-peaked [O I] and [Ca II]), 2008im (single-peaked [O I] and horned-like [Ca II]), 2005nb (single-peaked [O I] and flat-topped [Ca II]), 2004ao (horned-like [O I] and single-peaked [Ca II]). The central wavelengths of the typical nebular emissions are labeled by the vertical lines.

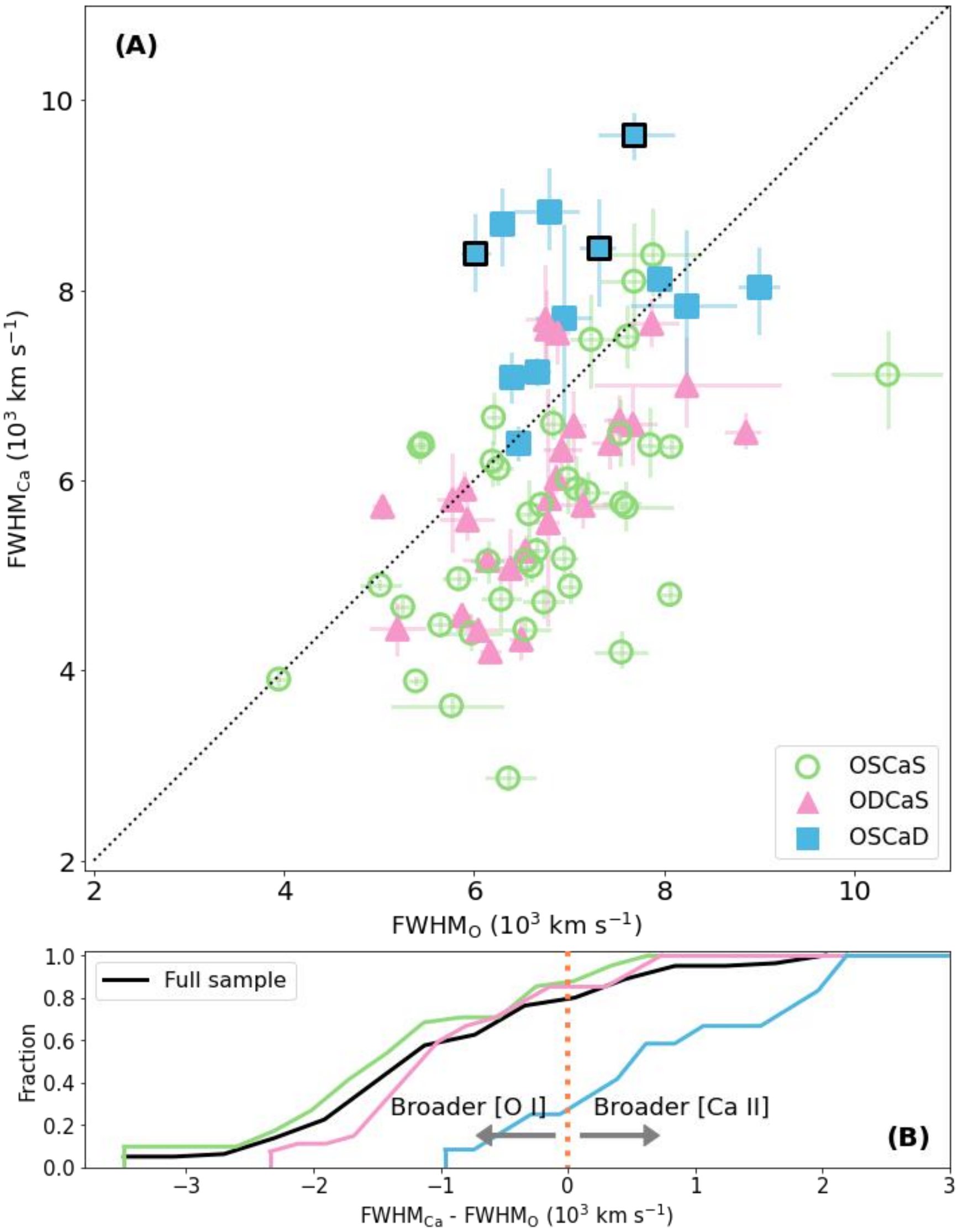

**Fig. 2 The analysis of the line widths. (A) The relation between the FWHMs of [O I] and [Ca II].** Objects in the different line profile categories (classified by the profiles of [O I] and [Ca II]) are labeled by different colors and markers (OSCaS: both [O I] and [Ca II] are single-peaked; ODCaS: [O I] is double-peaked and [Ca II] is single-peaked; OSCaD: [O I] is single-peaked and [Ca II] is double-peaked). Objects with the horned [Ca II] profiles are highlighted by the thick edges. Data are presented as mean value with error bar indicating the $16^{th}$ and $84^{th}$ percentiles. **(B) The statistics of the difference in line widths.** The cumulative distributions of the difference in [Ca II] and [O I] widths for objects in different line profile categories are compared.

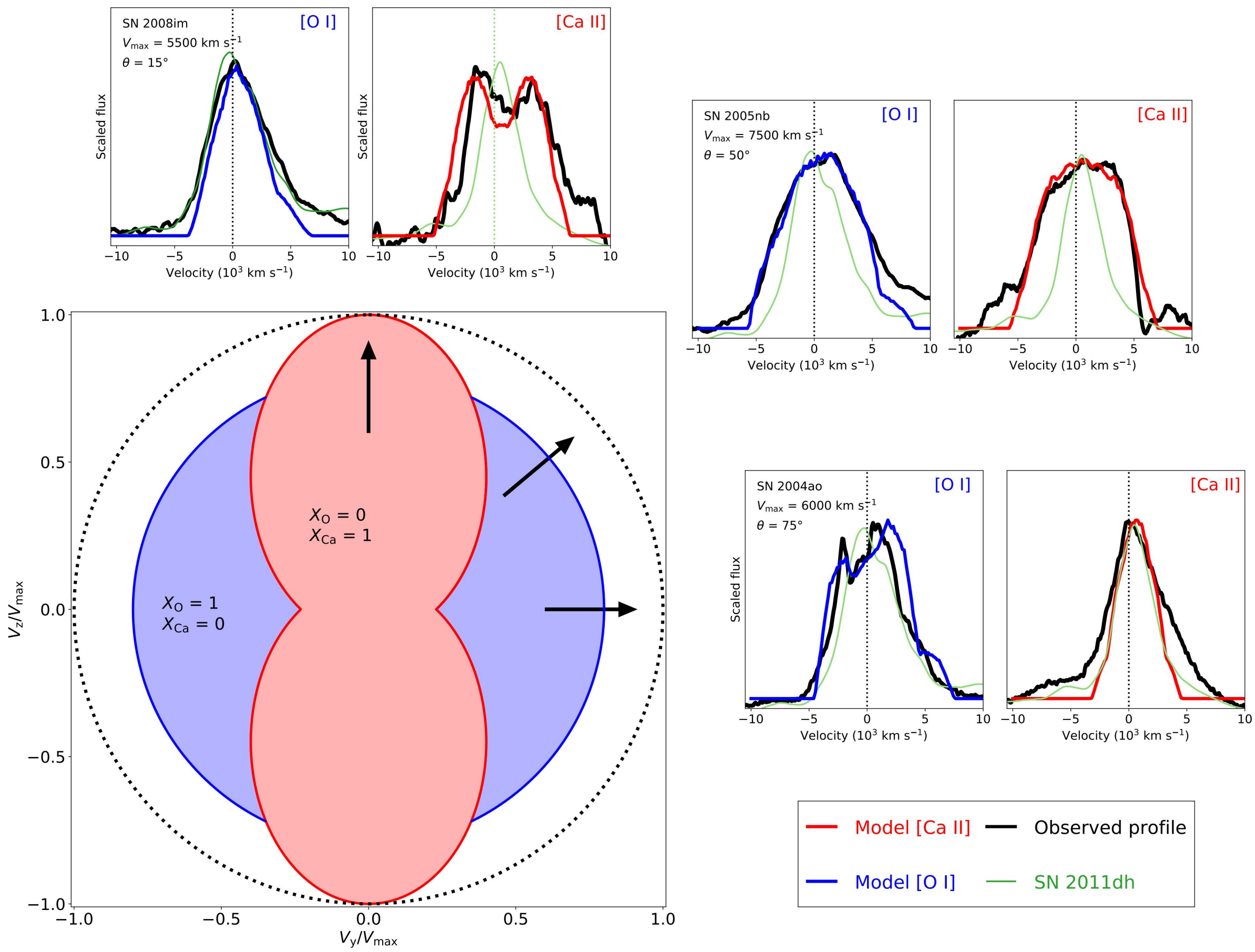

**Fig. 3 The configuration of the idealized axisymmetric model.** The O-rich region (labeled by blue) and the iron/calcium-rich ash (labeled by red) are complementary in their distributions. The profiles of the model [O I] (blue) and [Ca II] (red) strongly depend on the viewing angle, which are compared with the observed [O I] and [Ca II] profiles (black lines) of SN 2008im, SN 2005nb and SN 2004ao. In these panels, the [O I] and [Ca II] profiles of SN 2011dh, an OSCaS object, are plotted to illustrate the difference between the spherical and bipolar explosions.

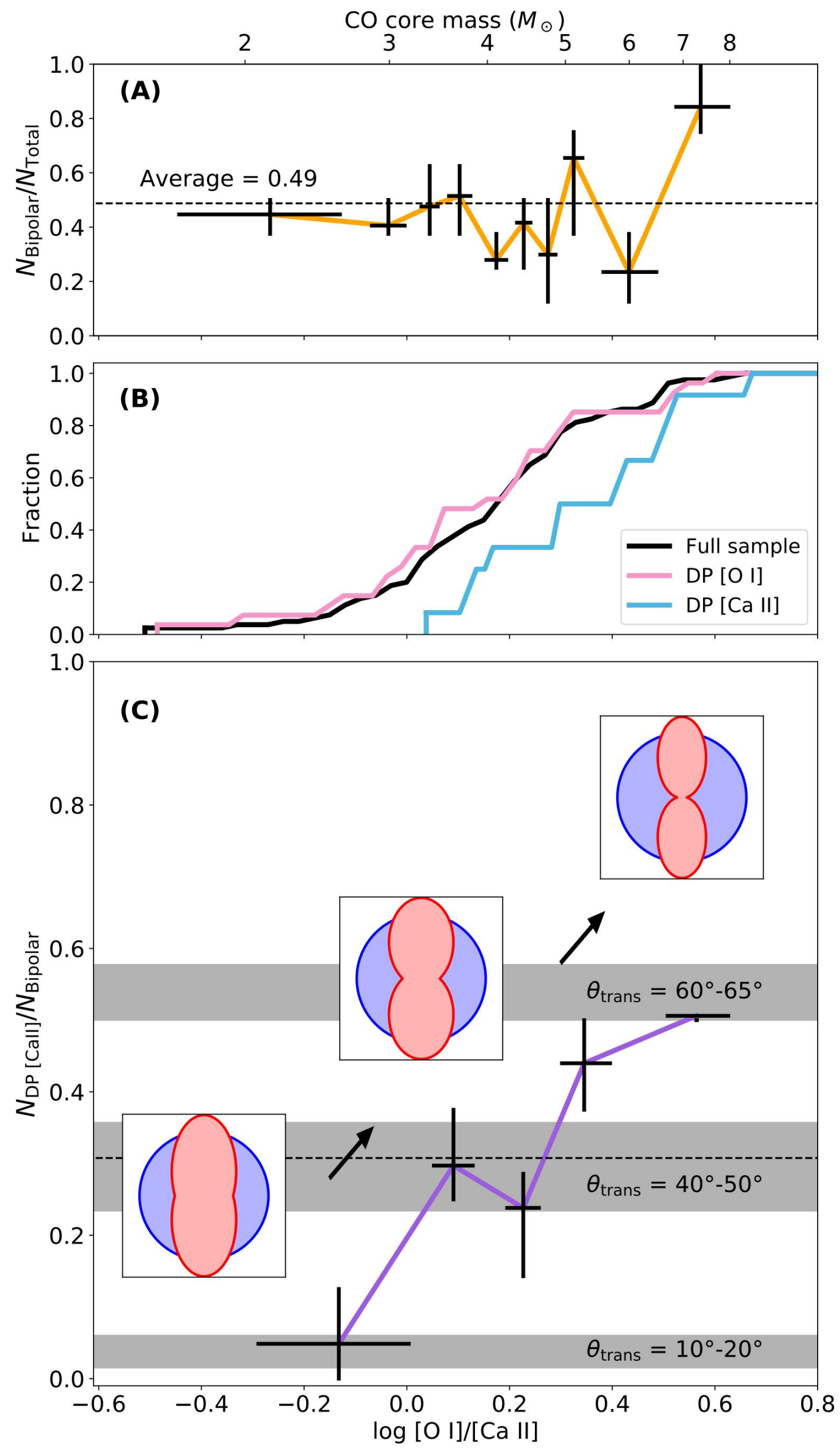

**Fig. 4 The relation between the explosion geometry and the CO mass of the progenitor. (A) The occurrence rate of bipolar SNe as a function of [O I]/[Ca II].** The bipolar rate of the full sample (0.49) is indicated by the dashed line. **(B) The cumulative distribution of the [O I]/[Ca II] of different line profile categories.** The ratio of the full sample (black), double-peaked [O I] (pink) and double-peaked [Ca II] (light blue) are compared; **(C) The occurrence rate of double-peaked [Ca II] as function of [O I]/[Ca II].** The insert panels schematically illustrate a sequence of the explosion geometry as a function of the CO core mass, which explains the observed relation. The three models have different transition angels $\theta_{trans}$ (when viewed above this angle, the [Ca II] profile transforms from double-peaked to single-peaked), and the shaded regions mark the expected double-peaked [Ca II] rates of these models assuming uniform distribution of solid angle. In panels (A) and (C), data are presented as mean value with error bar indicating the 16$^{th}$ and 84$^{th}$ percentiles.

## References

1. Khokhlov, A.M., Höflich, P.A., Oran, E.S., et al., Jet-induced Explosions of Core Collapse Supernovae. *Astrophys. J. Lett.* **524**, L107-L110 (1997). doi:10.1086/312305
2. Wang, L., Howell, D. A., Höflich, P., et al., Bipolar Supernova Explosions. *Astrophys. J. Lett.* **550**, 1030-1035 (2001). doi:10.1086/319822
3. Maeda, K. & Nomoto, K., Bipolar Supernova Explosions: Nucleosynthesis and Implications for Abundances in Extremely Metal-Poor Stars. *Astrophys. J.* **598**, 1163-1200 (2003). doi:10.1086/378948
4. Maeda, K., Nomoto, K., Mazzali, P.A. & Deng, J., Nebular Spectra of SN 1998bw Revisited: Detailed Study by One- and Two-dimensional Models. *Astrophys. J.* **640**, 854-877 (2006). doi:10.1086/500187
5. Woosley, S. E. & Bloom, J. S., The Supernova Gamma-Ray Burst Connection. *Annu. Rev. Astron. Astrophys.* **44**, 507-556 (2006). doi:10.1146/annurev.astro.43.072103.150558
6. Nomoto, K. I., Iwamoto, K., & Suzuki, T., The evolution and explosion of massive binary stars and Type Ib-Ic-IIb-IIL supernovae. *Phys. Rep.* **256**, 173-191 (1995). doi:10.1016/0370-1573(94)00107-E
7. Mazzali, P. A., Kawabata, K., Maeda, K. et al., An Asymmetric Energetic Type Ic Supernova Viewed Off-Axis, and a Link to Gamma Ray Bursts. *Sci.* **308**, 1284-1287 (2005). doi:10.1126/science.1111384
8. Maeda, K., Nakamura, T., Nomoto, K., et al., Explosive Nucleosynthesis in Aspherical Hypernova Explosions and Late-Time Spectra of SN 1998bw. *Astrophys. J.* **565**, 405-412 (2002). doi:10.1086/324487
9. Maeda, K., Kawabata, K., Mazzali, P. A., et al., Asphericity in Supernova Explosions from Late-Time Spectroscopy. *Sci.* **319**, 1220- (2008). doi:10.1126/science.1149437
10. Modjaz, M., Kirshner, R. P., Blondin, S., et al., Double-Peaked Oxygen Lines Are Not Rare in Nebular Spectra of Core-Collapse Supernovae. *Astrophys. J. Lett.* **687**, L9 (2008). doi:10.1086/593135
11. Taubenberger, S., Valenti, S., Benetti, S., et al., Nebular emission-line profiles of Type Ib/c supernovae - probing the ejecta asphericity. *Mon. Not. R. Astron. Soc.* **397**, 677-694 (2009). doi:10.1111/j.1365-2966.2009.15003.x
12. Milisavljevic, D., Fesen, R. A., Gerardy, C. L., et al., Doublets and Double Peaks: Late-Time [O I] λλ6300, 6364 Line Profiles of Stripped-Envelope, Core-Collapse Supernovae. *Astrophys. J.* **709**, 1343-1355 (2010). doi:10.1088/0004-637X/709/2/1343
13. Fang, Q., Maeda, K., Kuncarayakti, H., et al., Statistical Properties of the Nebular Spectra of 103 Stripped-envelope Core-collapse Supernovae. *Astrophys. J.* **928**, 151 (2022). doi:10.3847/1538-4357/ac4f60
14. Jerkstrand, A. "Spectra of Supernovae in the Nebular Phase." in *The Handbook of Supernovae.* 795 (Springer International Publishing AG, 2017)
15. Dessart, L., Hillier, D. J., Sukhbold, T., et al., Nebular phase properties of supernova Ibc from He-star explosions. *Astron. Astrophys.* **656**, A61 (2021). doi:10.1051/0004-6361/202141927
16. Prentice, S. J., Maguire, K., Siebenaler, L., et al., Oxygen and calcium nebular emission line relationships in core-collapse supernovae and Ca-rich transients. *Mon. Not. R. Astron. Soc.* **514**, 5686-5705 (2022). doi:10.1093/mnras/stac1657

17. van Baal, B. F. A., Jerkstrand, A., Wongwathanarat, A., et al., Modelling supernova nebular lines in 3D with EXTRASS. *Mon. Not. R. Astron. Soc.* **523,** 954 (2023). doi:10.1093/mnras/stad1488
18. Maurer, J. I., Mazzali, P. A., Deng, J., et al., Characteristic velocities of stripped-envelope core-collapse supernova cores. *Mon. Not. R. Astron. Soc.* **402**, 161-172 (2010). doi:10.1111/j.1365-2966.2009.15905.x
19. Fransson, C., & Chevalier, R. A., Late emission from supernovae - A window on stellar nucleosynthesis. *Astrophys. J.* **343**, 323 (1989). doi:10.1086/167707
20. Jerkstrand, A., Ergon, M., Smartt, S. J., et al., Late-time spectral line formation in Type IIb supernovae, with application to SN 1993J, SN 2008ax, and SN 2011dh. *Astron. Astrophys*. **573**, A12 (2015). doi:10.1051/0004-6361/201423983
21. Kuncarayakti, H., Maeda, K., Bersten, M. C., et al., Nebular phase observations of the Type-Ib supernova iPTF13bvn favour a binary progenitor. *Astron. Astrophys.* **579**, A95 (2015). doi:10.1051/0004-6361/201425604
22. Fang, Q., Maeda, K., Kuncarayakti, H., et al., A hybrid envelope-stripping mechanism for massive stars from supernova nebular spectroscopy. *Nature. Astron.* **3**, 434-439 (2019). doi:10.1038/s41550-019-0710-6
23. Dessart, L., Hillier, D. J., Woosley, S. E., et al., Modeling of the nebular-phase spectral evolution of stripped-envelope supernovae. New grids from 100 to 450 days. *Astron. Astrophys.* **677**, A7 (2023). doi:10.1051/0004-6361/202346626
24. Fang, Q. & Maeda, K., Inferring the Progenitor Mass-Kinetic Energy Relation of Stripped-envelope Core-collapse Supernovae from Nebular Spectroscopy. *Astrophys. J.* **949,** 93 (2023). doi:10.3847/1538-4357/acc5e7
25. Nakamura, K., Takiwaki, T., Kuroda, T., et al., Systematic features of axisymmetric neutrino-driven core-collapse supernova models in multiple progenitors. *Publ. Astron. Soc. Japan.* **67**, 107 (2015). doi:10.1093/pasj/psv073
26. Woosley, S. E., Heger, A., & Weaver, T. A., The evolution and explosion of massive stars. *Rev. Mod. Phys.* **74**, 1015-1071 (2002). doi:10.1103/RevModPhys.74.1015
27. Limongi, M. & Chieffi, A., Evolution, Explosion, and Nucleosynthesis of Core-Collapse Supernovae. *Astrophys. J.* **592**, 404-433 (2003). doi:10.1086/375703
28. Benjamin, D. J., Berger, J. O., Johannesson, M. et al. Redefine statistical significance., *Nat. Hum. Behav.* ***2,*** *6–10 (2018)*. doi: https://doi.org/10.1038/s41562-017-0189-z
29. Burrows, A. & Vartanyan, D., Core-collapse supernova explosion theory. *Nature.* **589**, 29-39 (2021). doi:10.1038/s41586-020-03059-w
30. Blondin, J. M., Mezzacappa, A., & DeMarino, C., Stability of Standing Accretion Shocks, with an Eye toward Core-Collapse Supernovae. *Astrophys. J.* **584,** 971 (2003). doi:10.1086/345812
31. Fernández, R., Three-dimensional simulations of SASI- and convection-dominated core-collapse supernovae. *Mon. Not. R. Astron. Soc.* **452,** 2071 (2015). doi:10.1093/mnras/stv1463
32. Matsumoto, J., Takiwaki, T., Kotake, K., et al., 2D numerical study for magnetic field dependence of neutrino-driven core-collapse supernova models. *Mon. Not. R. Astron. Soc.* **499**, 4174-4194 (2020). doi:10.1093/mnras/staa3095
33. Varma, V., Müller, B., & Schneider, F. R. N., 3D simulations of strongly magnetized non-rotating supernovae: explosion dynamics and remnant properties. *Mon. Not. R. Astron. Soc.* **518**, 3622-3636 (2023). doi:10.1093/mnras/stac3247
34. Reichert, M., Obergaulinger, M., Aloy, M. Á., et al., Magnetorotational supernovae: a

nucleosynthetic analysis of sophisticated 3D models. *Mon. Not. R. Astron. Soc.* **518**, 1557-1583 (2023). doi:10.1093/mnras/stac3185

35. Shivvers, I., Filippenko, A. V., Silverman, J. M., et al., The Berkeley sample of stripped-envelope supernovae. *Mon. Not. R. Astron. Soc.* **482**, 1545-1556 (2019). doi:10.1093/mnras/sty2719
36. Yaron. O & Gal-Yam. A., WISeREP-An Interactive Supernova Data Repository. *Publ. Astron. Soc. Pac.* **124**, 668-681 (2012). doi:doi:10.1086/666656
37. Galama, T. J., Vreeswijk, P. M., van Paradijs, J., et al., An unusual supernova in the error box of the γ-ray burst of 25 April 1998. *Nature.*, **395**, 670-672 (1998). doi:10.1038/27150
38. Iwamoto, K., Mazzali, P. A., Nomoto, K., et al., A hypernova model for the supernova associated with the γ-ray burst of 25 April 1998 *Nature.* **395**, 672-674 (1998). doi:10.1038/27155
39. Pian, E., Mazzali, P. A., Masetti, N., et al., An optical supernova associated with the X-ray flash XRF 060218. *Nature.* **442**, 1011-1013 (2006). doi:10.1038/nature05082
40. Mazzali, P. A., Deng, J., Nomoto, K., et al., A neutron-star-driven X-ray flash associated with supernova SN 2006aj. *Nature.* **442**, 1018-1020 (2006). doi:10.1038/nature05081
41. Milisavljevic, D., Margutti, R., Parrent, J. T., et al., The Broad-lined Type Ic SN 2012ap and the Nature of Relativistic Supernovae Lacking a Gamma-Ray Burst Detection. *Astrophys. J.* **799**, 51 (2015). doi:10.1088/0004-637X/799/1/51
42. Chakraborti, S., Soderberg, A., Chomiuk, L., et al., A Missing-link in the Supernova-GRB Connection: The Case of SN 2012ap. *Astrophys. J.* **805**, 187 (2015). doi:10.1088/0004-637X/805/2/187
43. Maeda, K., Tanaka, M., Nomoto, K., et al., The Unique Type Ib Supernova 2005bf at Nebular Phases: A Possible Birth Event of a Strongly Magnetized Neutron Star. *Astrophys. J.* **666**, 1069-1082 (2007). doi:10.1086/520054
44. Makarov, D., Prugniel, P., Terekhova, N., et al., HyperLEDA. III. The catalogue of extragalactic distances. *Astron. Astrophys.* **570**, A13 (2014). doi:10.1051/0004-6361/201423496
45. Fang, Q. & Maeda, K., The Origin of the Ha-like Structure in Nebular Spectra of Type IIb Supernovae. *Astrophys. J.* **864**, 47 (2018). doi:10.3847/1538-4357/aad096
46. Jerkstrand, A., Timmes, F. X., Magkotsios, G., et al., Constraints on Explosive Silicon Burning in Core-collapse Supernovae from Measured Ni/Fe Ratios. *Astrophys. J.* **807**, 110 (2015). doi:10.1088/0004-637X/807/1/110
47. Jerkstrand, A., Smartt, S. J., Sollerman, J., et al., Supersolar Ni/Fe production in the Type IIP SN 2012ec. *Mon. Not. R. Astron. Soc.* **448**, 2482-2494 (2015). doi:10.1093/mnras/stv087
48. Dong, S., Katz, B., Kushnir, D., et al., Type Ia supernovae with bimodal explosions are common - possible smoking gun for direct collisions of white dwarfs. *Mon. Not. R. Astron. Soc.* **454**, L61 (2015). doi:10.1093/mnrasl/slv129
49. Patat, F., Cappellaro, E., Danziger, J., et al., *Astrophys. The Metamorphosis of SN 1998bw. J.* **555,** 900 (2001). doi:10.1086/321526
50. Maeda, K., Kawabata, K., Tanaka, M., et al., SN 2006aj Associated with XRF 060218 at Late Phases: Nucleosynthesis Signature of a Neutron Star-driven Explosion. *Astrophys. J. Lett.* **658**, L5-L8 (2007). doi:10.1086/513564

# Supplementary information for “An aspherical distribution for the explosive burning ash of core-collapse supernovae”

Table of contents:

| SN name | Date | Line profile | FWHM ($10^3$ km s$^{-1}$) | | References |
|---|---|---|---|---|---|
| | | | [O I] | [Ca II] | |
| 1985F | 1985/04/01 | OSCaS | $5.24^{+0.06}_{-0.07}$ | $4.66^{+0.07}_{-0.06}$ | *S1* |
| 1987K | 1998/02/24 | OSCaS | $7.87^{+0.49}_{-0.49}$ | $8.37^{+0.46}_{-0.40}$ | *S2* |
| 1987M | 1988/02/25 | OSCaS | $10.34^{+0.56}_{-0.57}$ | $7.11^{+0.43}_{-0.55}$ | *S3* |
| 1990U | 1991/01/06 | ODCaS | $5.86^{+0.08}_{-0.07}$ | $4.58^{+0.11}_{-0.11}$ | *11* |
| 1990W | 1991/02/21 | OSCaS | $6.25^{+0.13}_{-0.13}$ | $6.12^{+0.14}_{-0.16}$ | *11* |
| 1990aj | 1991/01/29 | OSCaD | $6.78^{+0.29}_{-0.35}$ | $8.83^{+0.43}_{-0.38}$ | *11* |
| 1991A | 1991/04/07 | OSCaS | $7.67^{+0.34}_{-0.33}$ | $8.09^{+0.58}_{-0.71}$ | *S4* |
| 1993J | 1993/11/07 | OSCaS | $6.73^{+0.19}_{-0.20}$ | $4.71^{+0.15}_{-0.15}$ | *S5* |
| 1994I | 1994/09/02 | OSCaS | $7.60^{+0.08}_{-0.12}$ | $7.51^{+0.31}_{-0.31}$ | *S6* |
| 1996aq | 1997/04/02 | OSCaD | $7.68^{+0.40}_{-0.36}$ | $9.63^{+0.21}_{-0.23}$ | *11* |
| 1996cb | 1997/07/01 | OSCaS | $6.65^{+0.12}_{-0.13}$ | $5.25^{+0.13}_{-0.13}$ | *S7* |
| 1997X | 1997/05/13 | OSCaS | $7.22^{+0.21}_{-0.20}$ | $7.48^{+0.45}_{-0.48}$ | *11* |
| 1997dq | 1998/05/30 | OSCaS | $7.07^{+0.12}_{-0.12}$ | $5.91^{+0.32}_{-0.29}$ | *11* |
| 2000ew | 2001/03/17 | ODCaS | $6.91^{+0.19}_{-0.16}$ | $6.32^{+0.32}_{-0.30}$ | *11* |
| 2001ig | 2002/11/08 | ODCaS | $6.48^{+0.08}_{-0.09}$ | $4.32^{+0.19}_{-0.19}$ | *S8* |
| 2002ap | 2002/08/09 | OSCaD | $7.94^{+0.04}_{-0.04}$ | $8.11^{+0.17}_{-0.17}$ | *S9* |
| 2003bg | 2003/11/29 | OSCaS | $6.82^{+0.10}_{-0.11}$ | $6.59^{+0.17}_{-0.14}$ | *S10* |
| 2003gf | 2003/11/29 | OSCaS | $8.06^{+0.05}_{-0.05}$ | $6.35^{+0.09}_{-0.09}$ | *35* |
| 2004ao | 2004/11/14 | ODCaS | $6.52^{+0.10}_{-0.09}$ | $5.27^{+0.08}_{-0.08}$ | *S6* |
| 2004aw | 2004/11/14 | OSCaS | $6.28^{+0.18}_{-0.19}$ | $4.74^{+0.33}_{-0.28}$ | *S6* |
| 2004dk | 2005/05/11 | OSCaS | $8.05^{+0.08}_{-0.06}$ | $4.80^{+0.07}_{-0.07}$ | *S6* |
| 2004fe | 2005/07/06 | ODCaS | $8.22^{+0.97}_{-0.93}$ | $7.00^{+0.47}_{-0.42}$ | *9* |
| 2004gk | 2005/07/10 | OSCaS | $6.54^{+0.12}_{-0.12}$ | $5.15^{+0.24}_{-0.25}$ | *9* |
| 2004gn | 2005/07/06 | OSCaS | $6.53^{+0.25}_{-0.26}$ | $4.43^{+0.12}_{-0.12}$ | *9* |
| 2004gq | 2005/08/26 | OSCaS | $6.20^{+0.07}_{-0.05}$ | $6.65^{+0.24}_{-0.26}$ | *S6* |
| 2004gt | 2005/05/24 | ODCaS | $7.42^{+0.15}_{-0.17}$ | $6.39^{+0.23}_{-0.25}$ | *11* |
| 2004gv | 2005/08/26 | OSCaS | $6.57^{+0.12}_{-0.12}$ | $5.64^{+0.34}_{-0.38}$ | *9* |
| 2005N | 2005/01/22 | OSCaS | $7.65^{+0.25}_{-0.22}$ | $6.59^{+0.36}_{-0.40}$ | *S11* |
| 2005aj | 2005/08/25 | ODCaS | $6.78^{+0.04}_{-0.04}$ | $5.80^{+0.09}_{-0.09}$ | *13* |
| 2005bj | 2005/08/25 | OSCaS | $5.45^{+0.06}_{-0.07}$ | $6.38^{+0.10}_{-0.12}$ | *13* |
| 2005kl | 2006/06/30 | ODCaS | $6.86^{+0.07}_{-0.07}$ | $6.02^{+0.13}_{-0.14}$ | *9* |
| 2005kz | 2006/06/30 | OSCaD | $6.39^{+0.06}_{-0.06}$ | $7.08^{+0.23}_{-0.25}$ | *9* |
| 2005nb | 2006/06/30 | OSCaD | $8.99^{+0.18}_{-0.20}$ | $8.03^{+0.40}_{-0.47}$ | *9* |
| 2006F | 2006/06/30 | OSCaS | $5.42^{+0.13}_{-0.12}$ | $6.35^{+0.15}_{-0.15}$ | *9* |
| 2006G | 2006/06/30 | ODCaS | $6.13^{+0.29}_{-0.24}$ | $5.16^{+0.07}_{-0.08}$ | *13* |
| 2006T | 2006/11/26 | ODCaS | $5.89^{+0.06}_{-0.06}$ | $5.91^{+0.15}_{-0.11}$ | *9* |
| 2006ck | 2007/01/24 | OSCaD | $8.22^{+0.50}_{-0.56}$ | $7.83^{+0.79}_{-0.80}$ | *9* |
| 2006ep | 2006/12/24 | OSCaB | $6.65^{+0.12}_{-0.10}$ | $7.14^{+0.13}_{-0.12}$ | *13* |
| 2006gi | 2007/02/10 | OSCaS | $6.93^{+0.12}_{-0.10}$ | $5.17^{+0.21}_{-0.21}$ | *11* |
| 2007C | 2007/06/20 | OSCaS | $7.55^{+0.18}_{-0.18}$ | $5.76^{+0.12}_{-0.11}$ | *11* |
| 2007I | 2007/07/15 | ODCaS | $7.04^{+0.11}_{-0.11}$ | $6.57^{+0.33}_{-0.32}$ | *11* |
| 2007Y | 2007/09/22 | OSCaS | $5.83^{+0.16}_{-0.15}$ | $4.96^{+0.08}_{-0.08}$ | *S12* |
| 2007gr | 2008/02/12 | OSCaS | $6.80^{+0.07}_{-0.07}$ | $5.74^{+0.10}_{-0.08}$ | *35* |
| 2007uy | 2008/06/06 | OSCaD | $7.30^{+0.15}_{-0.17}$ | $8.44^{+0.50}_{-0.59}$ | *S6* |
| 2008D | 2008/06/07 | ODCaS | $7.51^{+0.14}_{-0.13}$ | $6.64^{+0.31}_{-0.29}$ | *S13* |
| 2008aq | 2008/06/26 | ODCaS | $5.02^{+0.07}_{-0.08}$ | $5.72^{+0.11}_{-0.11}$ | *S6* |
| 2008ax | 2008/11/24 | ODCaS | $6.15^{+0.10}_{-0.08}$ | $4.19^{+0.06}_{-0.06}$ | *S14* |
| 2008bo | 2008/10/27 | OSCaS | $5.96^{+0.31}_{-0.29}$ | $4.39^{+0.18}_{-0.16}$ | *35* |
| 2008fo | 2009/04/05 | OSCaS | $7.84^{+0.12}_{-0.11}$ | $6.37^{+0.37}_{-0.31}$ | *13* |
| 2008hh | 2009/08/18 | OSCaB | $6.94^{+0.27}_{-0.28}$ | $7.71^{+0.95}_{-1.15}$ | *13* |
| 2008ie | 2009/10/27 | ODCaS | $5.18^{+0.27}_{-0.27}$ | $4.45^{+0.26}_{-0.26}$ | *13* |
| 2008im | 2009/08/18 | OSCaD | $6.00^{+0.13}_{-0.12}$ | $8.38^{+0.40}_{-0.37}$ | *13* |
| 2009C | 2009/10/26 | ODCaS | $5.77^{+0.12}_{-0.14}$ | $5.79^{+0.46}_{-0.53}$ | *13* |
| 2009K | 2009/10/26 | OSCaS | $6.18^{+0.03}_{-0.03}$ | $6.20^{+0.18}_{-0.24}$ | *13* |
| 2009jf | 2010/06/19 | OSCaS | $7.19^{+0.20}_{-0.23}$ | $5.87^{+0.20}_{-0.18}$ | *S15* |
| 2009jy | 2010/05/06 | OSCaS | $6.98^{+0.17}_{-0.16}$ | $6.01^{+0.42}_{-0.34}$ | *13* |
| 2009ka | 2010/05/06 | OSCaS | $3.94^{+0.08}_{-0.08}$ | $3.90^{+0.07}_{-0.08}$ | *13* |
| 2010as | 2010/08/05 | ODCaS | $6.87^{+0.10}_{-0.12}$ | $7.55^{+0.27}_{-0.30}$ | *S16* |
| 2010mb | 2011/03/04 | OSCaS | $6.35^{+0.26}_{-0.22}$ | $2.86^{+0.11}_{-0.10}$ | *S17* |
| 2011bm | 2012/01/22 | ODCaS | $7.85^{+0.26}_{-0.20}$ | $7.65^{+0.24}_{-0.22}$ | *S18* |
| 2011dh | 2011/12/24 | OSCaS | $5.38^{+0.05}_{-0.05}$ | $3.88^{+0.02}_{-0.02}$ | *S19* |
| 2011ei | 2012/06/18 | ODCaS | $6.73^{+0.17}_{-0.17}$ | $7.69^{+0.55}_{-0.52}$ | *S20* |
| 2011fu | 2012/07/20 | OSCaS | $7.59^{+0.47}_{-0.39}$ | $5.72^{+0.24}_{-0.22}$ | *S21* |
| 2011hs | 2012/06/21 | OSCaS | $5.64^{+0.14}_{-0.13}$ | $4.47^{+0.08}_{-0.10}$ | *S22* |
| 2012P | 2012/08/08 | ODCaS | $7.14^{+0.27}_{-0.24}$ | $5.74^{+0.22}_{-0.22}$ | *S23* |
| 2012au | 2012/12/19 | OSCaS | $7.53^{+0.11}_{-0.10}$ | $6.49^{+0.33}_{-0.32}$ | *S24* |
| 2012dy | 2012/12/23 | ODCaS | $8.85^{+0.14}_{-0.20}$ | $6.51^{+0.17}_{-0.16}$ | *36* |
| 2012fh | 2012/11/14 | ODCaS | $6.75^{+0.12}_{-0.12}$ | $7.60^{+0.37}_{-0.45}$ | *35* |
| 2013ak | 2013/09/13 | OSCaS | $7.54^{+0.25}_{-0.25}$ | $4.19^{+0.20}_{-0.15}$ | *16* |
| 2013df | 2013/12/21 | OSCaS | $5.00^{+0.21}_{-0.19}$ | $4.89^{+0.03}_{-0.03}$ | *S25* |
| 2013ge | 2014/04/28 | OSCaS | $7.00^{+0.06}_{-0.05}$ | $4.88^{+0.14}_{-0.16}$ | *S26* |
| 2014C | 2014/08/25 | ODCaS | $6.03^{+0.18}_{-0.20}$ | $4.42^{+0.09}_{-0.08}$ | *35* |
| 2014eh | 2015/06/16 | OSCaS | $6.60^{+0.04}_{-0.04}$ | $5.10^{+0.14}_{-0.14}$ | *35* |
| 2014L | 2014/06/29 | OSCaS | $6.15^{+0.14}_{-0.16}$ | $5.15^{+0.19}_{-0.21}$ | *35* |
| 2015Q | 2016/01/07 | OSCaS | $5.75^{+0.52}_{-0.61}$ | $3.62^{+0.06}_{-0.06}$ | *35* |
| 2015ah | 2016/01/07 | OSCaD | $6.28^{+0.05}_{-0.05}$ | $8.70^{+0.35}_{-0.42}$ | *35* |
| ASASSN14az | 2014/11/25 | ODCaS | $6.77^{+0.09}_{-0.11}$ | $5.56^{+1.37}_{-1.08}$ | *35* |
| PTF12gzk | 2013/06/10 | OSCaD | $6.47^{+0.05}_{-0.04}$ | $6.39^{+0.16}_{-0.16}$ | *35* |
| iPTF13bvn | 2014/05/28 | ODCaS | $5.92^{+0.27}_{-0.26}$ | $5.59^{+0.21}_{-0.20}$ | *S23* |
| iPTF15dtg | 2016/10/31 | ODCaS | $6.37^{+0.17}_{-0.15}$ | $5.09^{+0.38}_{-0.37}$ | *S27* |

**Supplementary Tab. 1 SNe sample in this work.** The columns are (from left to right): name of the SN, the observation date of the nebular spectrum, the line profiles of [O I] and [Ca II], the FWHM of [O I], the FWHM of [Ca II] and references.

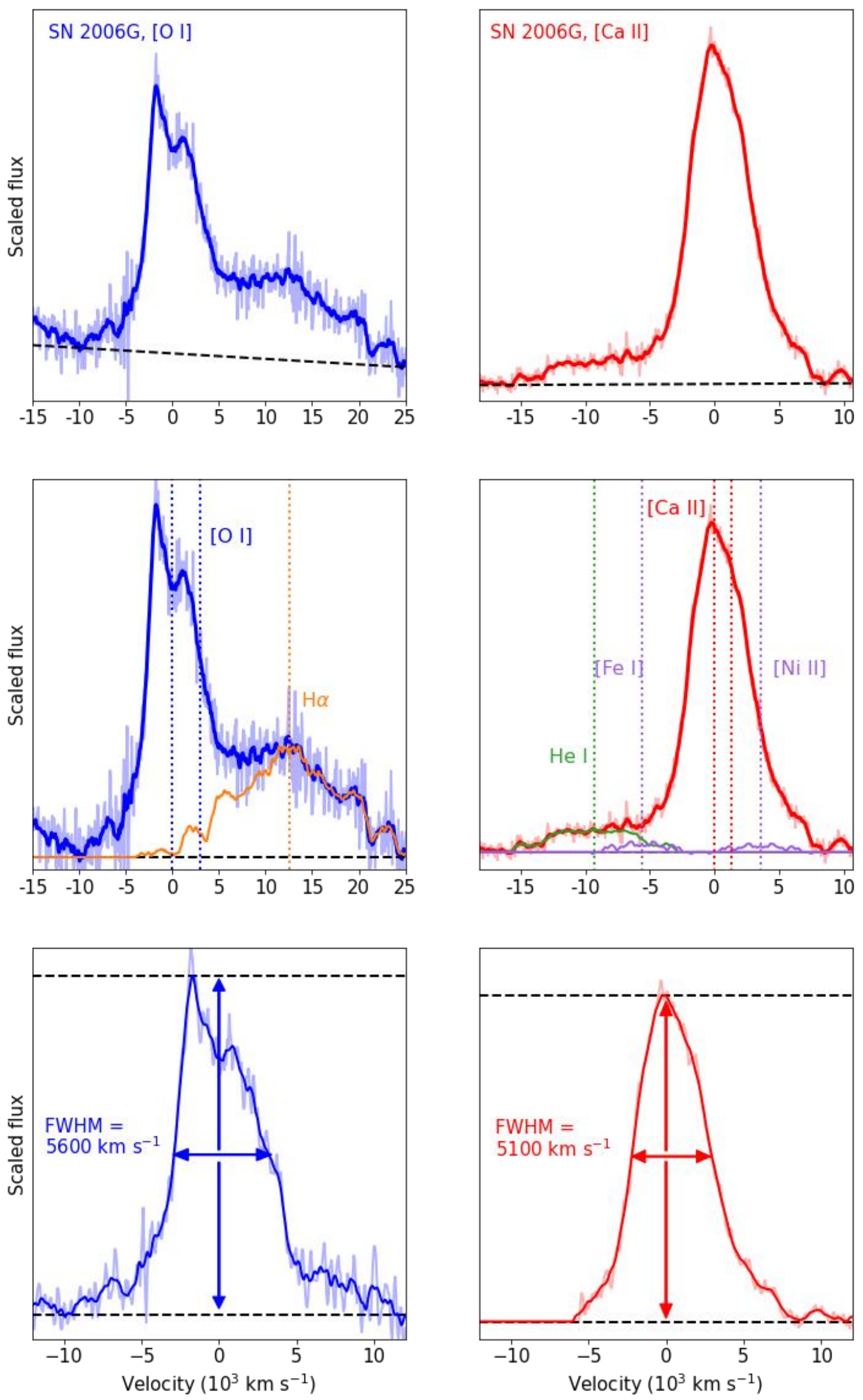

**Supplementary Fig. 1 A demonstration of the measurement procedures of line widths. Left panels:** measurement for [O I]. **Right panels:** measurement for [Ca II]. The continuum (black dashed line) is firstly excised (upper panels). Other species that contaminate [O I] and [Ca II], including Hα, He I, [Fe I] and [Ni II], are labeled by different colors (middle panels). The FWHM are measured after the continuum and the contamination are subtracted (lower panels).

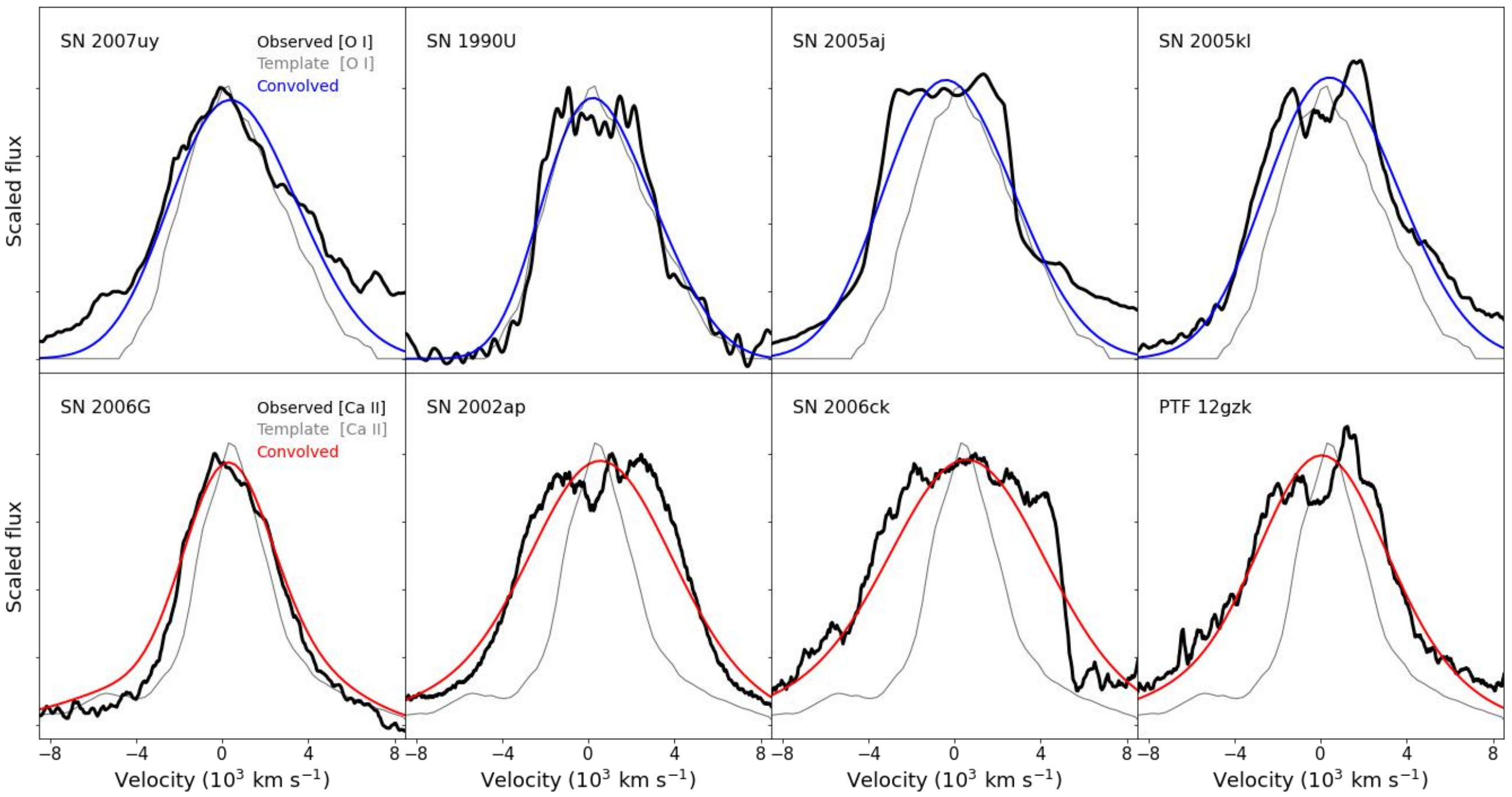

**Supplementary Fig. 2 A demonstration of some examples of the line profiles classifications.** The templates (grey lines) are convolved with Gaussian kernels to match with the widths of the observed profiles (black lines). The convolved templates are illustrated by the blue ([O I]) and red ([Ca II]) lines. The classifications of the profiles are (from left to right): single-peaked, flat-topped, flat-topped, horned-like. It should be noted that distinguishing between flat-topped and horn-like profiles can sometimes be ambiguous, as exemplified by SN 2002ap and PTF12gzk, while they both notably differ from the convolved SP templates.

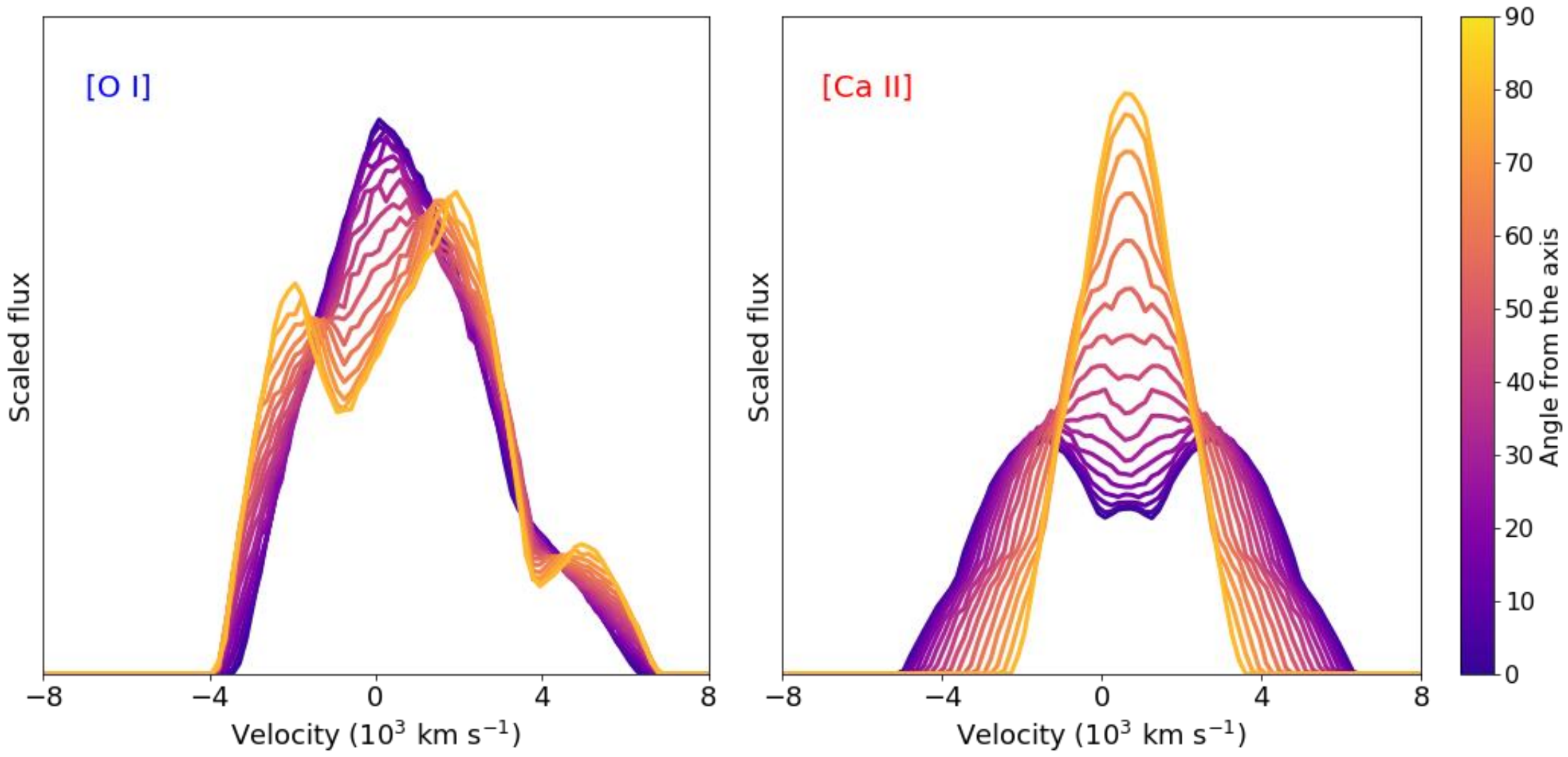

**Supplementary Fig. 3 The model profiles with different viewing angles. Left panel:** [O I] profiles. **Right panel:** [Ca II] profiles. The models are fixed to have $V_{\max}$ = 5000 km s$^{-1}$, and the profiles when viewed from different angles from the axis are labeled by different colors, with $\theta$ varying from 0º (blue end) to 90º (orange end).

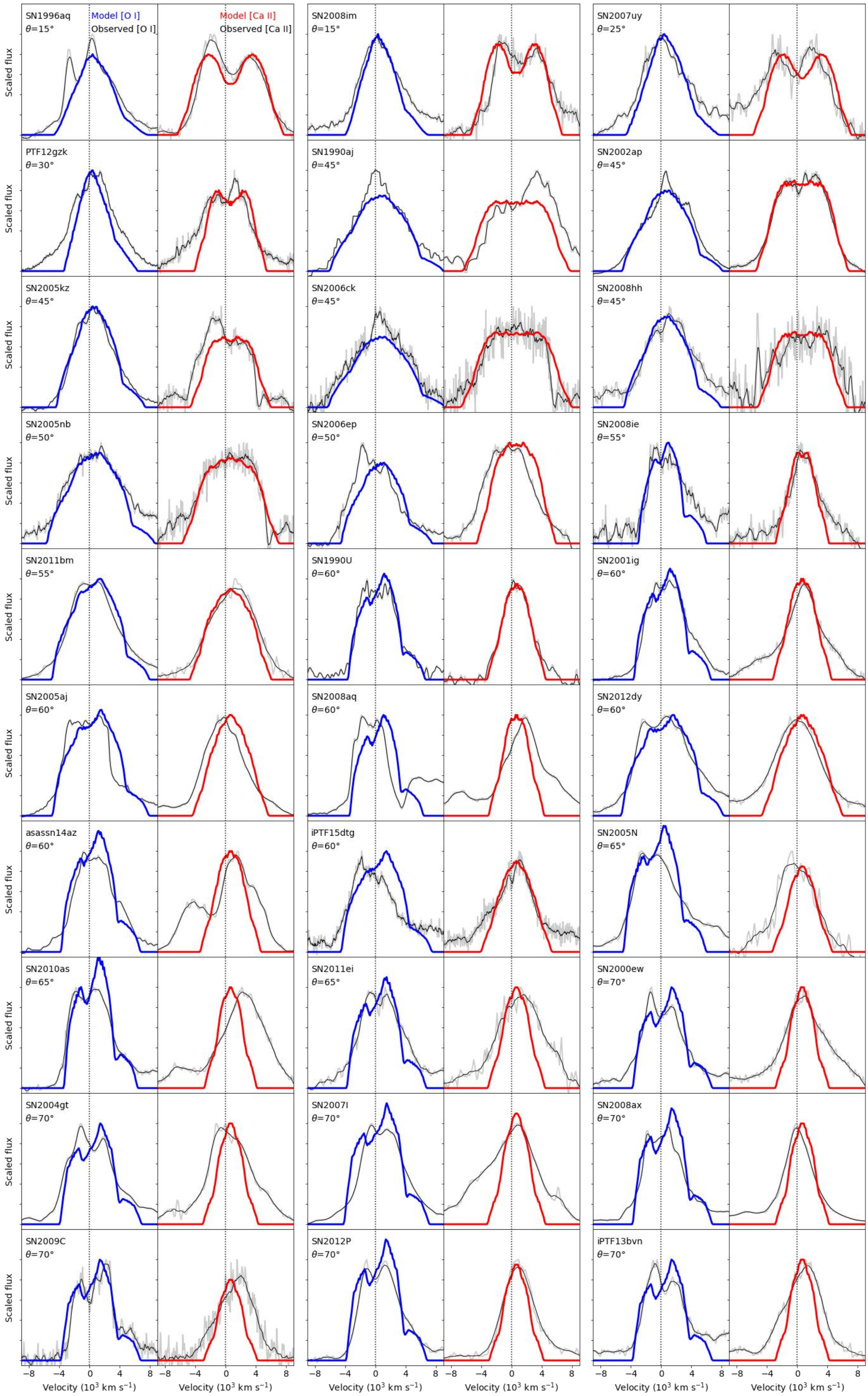

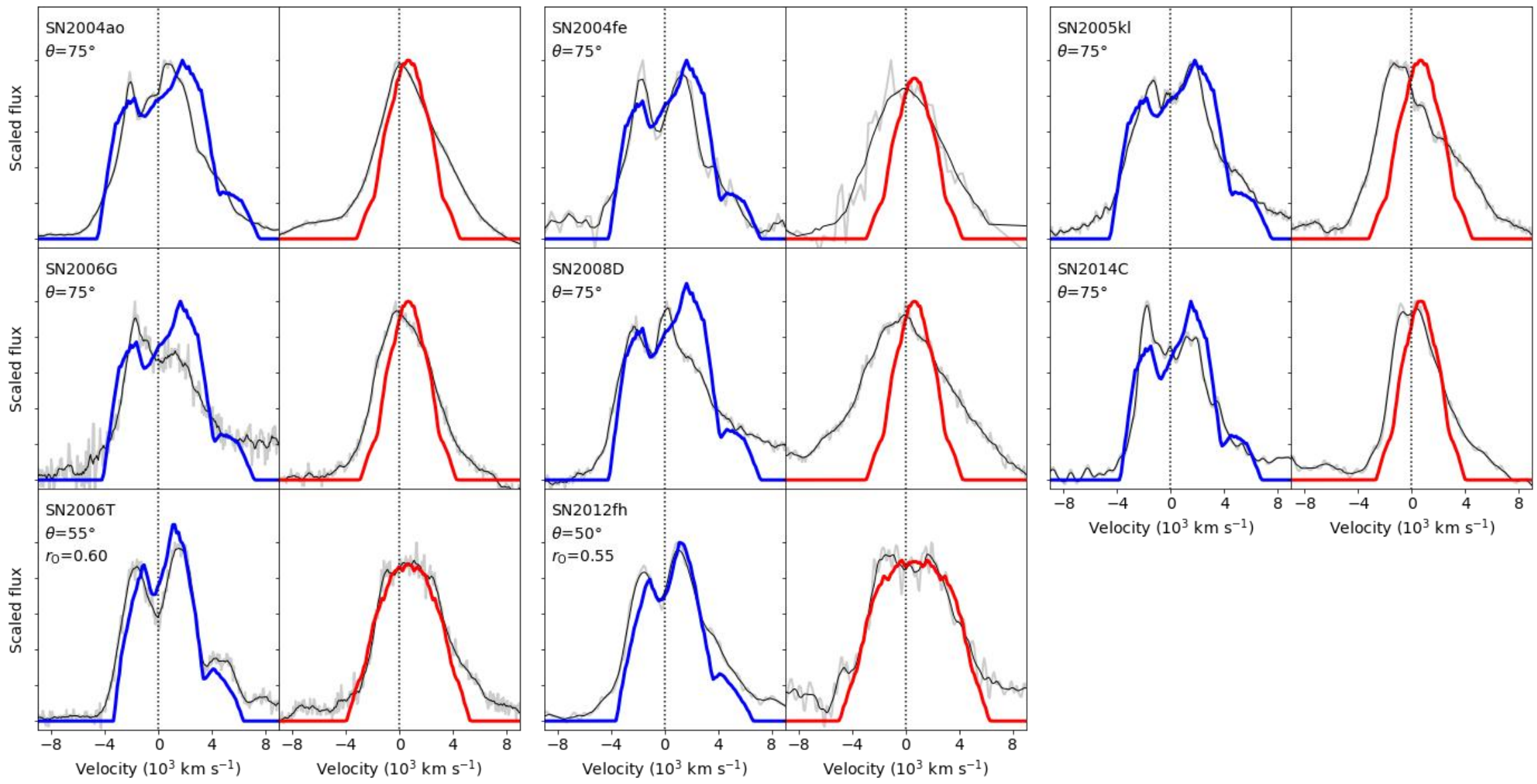

**Supplementary Fig. 4 The applications of the model to the observed line profiles.** The [O I] (blue) and [Ca II] (red) profiles synthesized from the axisymmetric model are compared with observed profiles (black) of the objects considered as bipolar explosion in this work.

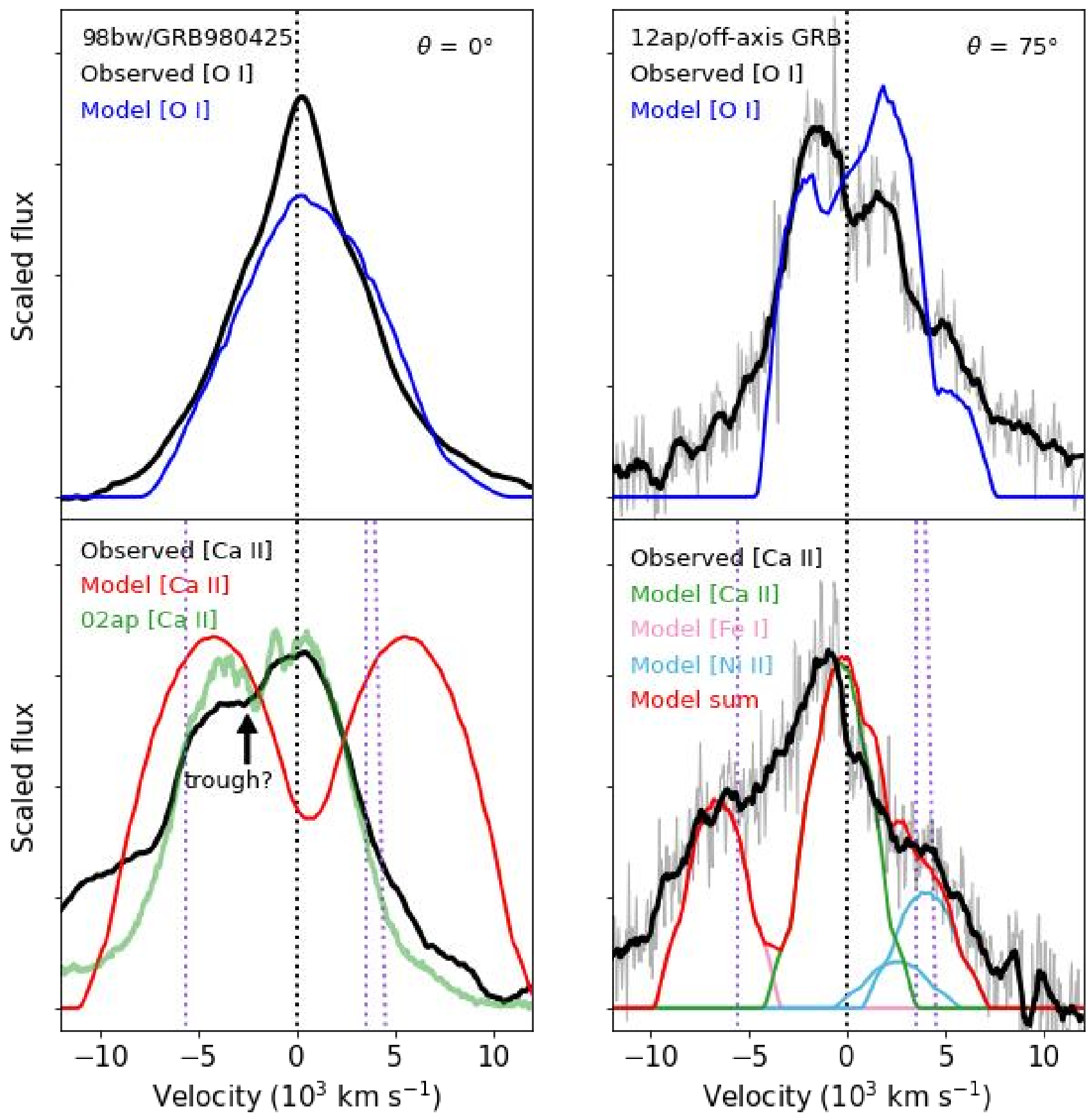

**Supplementary Fig. 5 The [O I] and [Ca II] profiles of SNe associated with GRBs. Left panels:** SN 1998bw (GRB-SNe viewed on-axis). **Right panels:** SN 2012ap (GRB-SNe viewed off-axis). The [O I] and [Ca II] profiles are shown in upper and lower panels. The model profiles, viewed from $\theta = 0°$ and $\theta = 75°$, are compared respectively. The black dotted line indicates 0 km s$^{-1}$, and the purple dotted lines indicate the central wavelengths of [Fe I] λ7155, [Ni II] λ7377 and [Ni II] λ7412. In the lower panels, the [Ca II] profile of SN 2002ap (green line), blueshifted by 2000 km s$^{-1}$, is plotted to compare with SN 1998bw. For SN 2012ap, the sum of the model emissions from the iron-peak elements (red), i.e., [Fe I] λ7155 (pink), [Ca II] λλ7291,7323 (green) and [Ni II] λ7377 (light blue) are plotted for comparison.

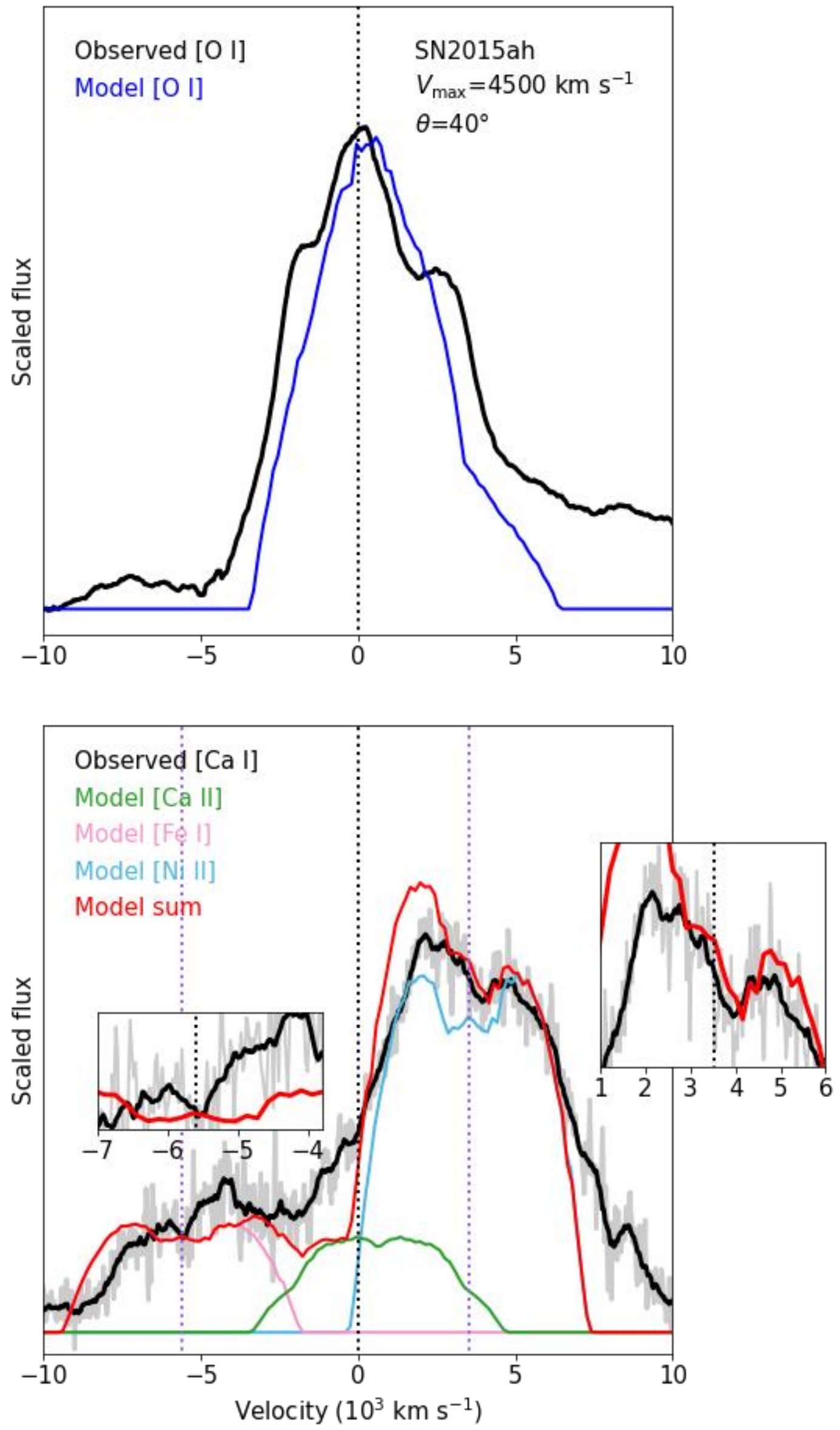

**Supplementary Fig. 6 The application of the model to SN 2015ah. Upper panel: [O I] profile. Lower panel: [Ca II] complex.** Assuming that Ni and Fe share the same spatial distribution as Ca, the sum of the model emissions from the iron-peak elements (red), i.e., [Fe I] λ7155 (pink), [Ca II] λλ7291,7323 (green) and [Ni II] λ7377 (light blue) produces the observed [Ca II] complex (black). The central wavelengths of [Fe I], [Ca II] and [Ni II] are labeled by the vertical dotted lines. At the same time, the model [O I] (blue line in the upper panel) also aligns with the observed single-peak profile.

## Supplementary References

S1. Filippenko, A. V. & Sargent, W. L. W., The unique supernova (1985f) in NGC 4618. *Astron. J.* **91**, 691-696 (1986). doi:10.1086/114051

S2. Filippenko, A. V., Supernova 1987K: Type II in Youth, Type Ib in Old Age. *Astron. J.* **96**, 1941 (1988). doi:10.1086/114940

S3. Filippenko, A. V., Porter, A. C., & Sargent, W. L. W., The Type IC (Helium-Poor Ib) Supernova 1987M: Transition to the Supernebular Phase. *Astron. J.* **100**, 1575 (1990). doi:10.1086/115618

S4. Matheson, T., Filippenko, A. V., Li, W., et al., Optical Spectroscopy of Type IB/C Supernovae. *Astron. J.* **121**, 1648-1675 (2001). doi:10.1086/319390

S5. Matheson, T., Filippenko, A. V., Ho, L. C., et al., Detailed Analysis of Early to Late-Time Spectra of Supernova 1993J. *Astron. J.* **120**, 1499-1515 (2000). doi:10.1086/301519

S6. Modjaz, M., Blondin, S., Kirshner, R. P., et al., Optical Spectra of 73 Stripped-envelope Core-collapse Supernovae. *Astron. J.* **147**, 99 (2014). doi:10.1088/0004-6256/147/5/99

S7. Qiu, Y., Li, W., Qiao, Q., et al., The Study of a Type IIB Supernova: SN 1996CB. *Astron. J.* **117**, 736-743 (1999). doi:10.1086/300731

S8. Silverman, J. M., Mazzali, P., Chornock, R., et al., Optical Spectroscopy of the Somewhat Peculiar Type IIb Supernova 2001ig. *Publ. Astron. Soc. Pac.* **121**, 689 (2009). doi:10.1086/603653

S9. Foley, R. J., Papenkova, M. S., Swift, B. J., et al., Optical Photometry and Spectroscopy of the SN 1998bw-like Type Ic Supernova 2002ap. *Publ. Astron. Soc. Pac.* **115**, 1220-1235 (2003). doi:10.1086/378242

S10. Hamuy, M., Deng, J., Mazzali, P. A., et al., Supernova 2003bg: The First Type IIb Hypernova. *Astrophys. J.* **703**, 1612-1623 (2009). doi:10.1088/0004-637X/703/2/1612

S11. Harutyunyan, A. H., Pfahler, P., Pastorello, A., et al., ESC supernova spectroscopy of non-ESC targets. *Astron. Astrophys.* **488**, 383 (2008). doi:10.1051/0004-6361:20078859

S12. Stritzinger, M., Mazzali, P., Phillips, M. M., et al., The He-Rich Core-Collapse Supernova 2007Y: Observations from X-Ray to Radio Wavelengths. *Astrophys. J.* **696**, 713-728 (2009). doi:10.1088/0004-637X/696/1/713

S13. Modjaz, M., Li, W., Butler, N., et al., From Shock Breakout to Peak and Beyond: Extensive Panchromatic Observations of the Type Ib Supernova 2008D Associated with Swift X-ray Transient 080109. *Astrophys. J.* **702**, 226-248 (2009). doi:10.1088/0004-637X/702/1/226

S14. Taubenberger, S., Navasardyan, H., Maurer, J. I., et al., The He-rich stripped-envelope core-collapse supernova 2008ax. *Mon. Not. R. Astron. Soc.* **413**, 2140-2156 (2011). doi:10.1111/j.1365-2966.2011.18287.x

S15. Sahu, D. K., Gurugubelli, U. K., Anupama, G. C., et al., Optical studies of SN 2009jf: a Type Ib supernova with an extremely slow decline and aspherical signature. *Mon. Not. R. Astron. Soc.* **413**, 2583-2594 (2011). doi:10.1111/j.1365-2966.2011.18326.x

S16. Folatelli, G., Bersten, M. C., Kuncarayakti, H., et al., Supernova 2010as: The Lowest-velocity Member of a Family of Flat-velocity Type IIb Supernovae. *Astrophys. J.* **792**, 7 (2014). doi:10.1088/0004-637X/792/1/7

S17. Ben-Ami, S., Gal-Yam, A., Mazzali, P. A., et al., SN 2010mb: Direct Evidence for a Supernova Interacting with a Large Amount of Hydrogen-free Circumstellar Material. *Astrophys. J.* **785**, 37 (2014). doi:10.1088/0004-637X/785/1/37

S18. Valenti, S., Taubenberger, S., Pastorello, A., et al., A Spectroscopically Normal Type Ic Supernova from a Very Massive Progenitor. *Astrophys. J. Lett.* **749**, L28 (2012). doi:10.1088/2041-8205/749/2/L28

S19. Shivvers, I., Mazzali, P., Silverman, J. M., et al., Nebular spectroscopy of the nearby Type IIb supernova 2011dh. *Mon. Not. R. Astron. Soc.* **436**, 3614-3625 (2013). doi:10.1093/mnras/stt1839

S20. Milisavljevic, D., Margutti, R., Soderberg, A. M., et al., Multi-wavelength Observations of Supernova 2011ei: Time-dependent Classification of Type IIb and Ib Supernovae and Implications for Their Progenitors. *Astrophys. J.* **767**, 71 (2013). doi:10.1088/0004-637X/767/1/71

S21. Morales-Garoffolo, A., Elias-Rosa, N., Bersten, M., et al., SN 2011fu: a type IIb supernova with a luminous double-peaked light curve. *Mon. Not. R. Astron. Soc.* **454**, 95-114 (2015). doi:10.1093/mnras/stv1972

S22. Bufano, F., Pignata, G., Bersten, M., et al., SN 2011hs: a fast and faint Type IIb supernova from a supergiant progenitor. *Mon. Not. R. Astron. Soc.* **439**, 1807-1828 (2014). doi:10.1093/mnras/stu065

S23. Fremling, C., Sollerman, J., Taddia, F., et al., PTF12os and iPTF13bvn. Two stripped-envelope supernovae from low-mass progenitors in NGC 5806. *Astron. Astrophys.* **593**, A68 (2016). doi:10.1051/0004-6361/201628275

S24. Milisavljevic, D., Soderberg, A. M., Margutti, R., et al., SN 2012au: A Golden Link between Superluminous Supernovae and Their Lower-luminosity Counterparts. *Astrophys. J. Lett.* **770**, L38 (2013). doi:10.1088/2041-8205/770/2/L38

S25. Maeda, K., Hattori, T., Milisavljevic, D., et al., Type IIb Supernova 2013df Entering into an Interaction Phase: A Link between the Progenitor and the Mass Loss. *Astrophys. J.* **807**, 35 (2015). doi:10.1088/0004-637X/807/1/35

S26. Drout, M. R., Milisavljevic, D., Parrent, J., et al., The Double-peaked SN 2013ge: A Type Ib/c SN with an Asymmetric Mass Ejection or an Extended Progenitor Envelope. *Astrophys. J.* **821**, 57 (2016). doi:10.3847/0004-637X/821/1/57

S27. Taddia, F., Sollerman, J., Fremling, C., et al., The luminous late-time emission of the type-Ic supernova iPTF15dtg - evidence for powering from a magnetar? *Astron. Astrophys.* **621**, A64 (2019). doi:10.1051/0004-6361/201833688